\documentclass[a4paper,11pt]{article}
\usepackage{jheppub} % for details on the use of the package, please
\usepackage[T1]{fontenc} % if needed
\usepackage{braket}
\usepackage{csquotes}
\usepackage{rotating}
\usepackage{graphicx}
\usepackage{relsize}
\usepackage{multirow}
\newcounter{AffiliationCounter}
\stepcounter{AffiliationCounter}\edef\instBilbao{\protect\theAffiliationCounter}
\stepcounter{AffiliationCounter}\edef\instBonn{\protect\theAffiliationCounter}
\stepcounter{AffiliationCounter}\edef\instBNL{\protect\theAffiliationCounter}
\stepcounter{AffiliationCounter}\edef\instBINP{\protect\theAffiliationCounter}
\stepcounter{AffiliationCounter}\edef\instCharles{\protect\theAffiliationCounter}
\stepcounter{AffiliationCounter}\edef\instChonnam{\protect\theAffiliationCounter}
\stepcounter{AffiliationCounter}\edef\instCincinnati{\protect\theAffiliationCounter}
\stepcounter{AffiliationCounter}\edef\instDESY{\protect\theAffiliationCounter}
\stepcounter{AffiliationCounter}\edef\instFlorida{\protect\theAffiliationCounter}
\stepcounter{AffiliationCounter}\edef\instFuJen{\protect\theAffiliationCounter}
\stepcounter{AffiliationCounter}\edef\instFudan{\protect\theAffiliationCounter}
\stepcounter{AffiliationCounter}\edef\instGoettingen{\protect\theAffiliationCounter}
\stepcounter{AffiliationCounter}\edef\instSokendai{\protect\theAffiliationCounter}
\stepcounter{AffiliationCounter}\edef\instGyeongsang{\protect\theAffiliationCounter}
\stepcounter{AffiliationCounter}\edef\instHanyang{\protect\theAffiliationCounter}
\stepcounter{AffiliationCounter}\edef\instHawaii{\protect\theAffiliationCounter}
\stepcounter{AffiliationCounter}\edef\instKEK{\protect\theAffiliationCounter}
\stepcounter{AffiliationCounter}\edef\instJPARC{\protect\theAffiliationCounter}
\stepcounter{AffiliationCounter}\edef\instHSE{\protect\theAffiliationCounter}
\stepcounter{AffiliationCounter}\edef\instJuelich{\protect\theAffiliationCounter}
\stepcounter{AffiliationCounter}\edef\instIKER{\protect\theAffiliationCounter}
\stepcounter{AffiliationCounter}\edef\instIISERM{\protect\theAffiliationCounter}
\stepcounter{AffiliationCounter}\edef\instIITB{\protect\theAffiliationCounter}
\stepcounter{AffiliationCounter}\edef\instIITH{\protect\theAffiliationCounter}
\stepcounter{AffiliationCounter}\edef\instIITM{\protect\theAffiliationCounter}
\stepcounter{AffiliationCounter}\edef\instIndiana{\protect\theAffiliationCounter}
\stepcounter{AffiliationCounter}\edef\instIHEP{\protect\theAffiliationCounter}
\stepcounter{AffiliationCounter}\edef\instProtvino{\protect\theAffiliationCounter}
\stepcounter{AffiliationCounter}\edef\instVienna{\protect\theAffiliationCounter}
\stepcounter{AffiliationCounter}\edef\instNapoli{\protect\theAffiliationCounter}
\stepcounter{AffiliationCounter}\edef\instTorino{\protect\theAffiliationCounter}
\stepcounter{AffiliationCounter}\edef\instJAEA{\protect\theAffiliationCounter}
\stepcounter{AffiliationCounter}\edef\instJSI{\protect\theAffiliationCounter}
\stepcounter{AffiliationCounter}\edef\instKarlsruhe{\protect\theAffiliationCounter}
\stepcounter{AffiliationCounter}\edef\instIPMU{\protect\theAffiliationCounter}
\stepcounter{AffiliationCounter}\edef\instKennesaw{\protect\theAffiliationCounter}
\stepcounter{AffiliationCounter}\edef\instKAU{\protect\theAffiliationCounter}
\stepcounter{AffiliationCounter}\edef\instKISTI{\protect\theAffiliationCounter}
\stepcounter{AffiliationCounter}\edef\instKorea{\protect\theAffiliationCounter}
\stepcounter{AffiliationCounter}\edef\instKyotoSangyo{\protect\theAffiliationCounter}
\stepcounter{AffiliationCounter}\edef\instKyoto{\protect\theAffiliationCounter}
\stepcounter{AffiliationCounter}\edef\instKyungpook{\protect\theAffiliationCounter}
\stepcounter{AffiliationCounter}\edef\instLAL{\protect\theAffiliationCounter}
\stepcounter{AffiliationCounter}\edef\instLebedev{\protect\theAffiliationCounter}
\stepcounter{AffiliationCounter}\edef\instLNNU{\protect\theAffiliationCounter}
\stepcounter{AffiliationCounter}\edef\instLjubljana{\protect\theAffiliationCounter}
\stepcounter{AffiliationCounter}\edef\instLMU{\protect\theAffiliationCounter}
\stepcounter{AffiliationCounter}\edef\instLuther{\protect\theAffiliationCounter}
\stepcounter{AffiliationCounter}\edef\instMNIT{\protect\theAffiliationCounter}
\stepcounter{AffiliationCounter}\edef\instMaribor{\protect\theAffiliationCounter}
\stepcounter{AffiliationCounter}\edef\instMPI{\protect\theAffiliationCounter}
\stepcounter{AffiliationCounter}\edef\instMelbourne{\protect\theAffiliationCounter}
\stepcounter{AffiliationCounter}\edef\instMississippi{\protect\theAffiliationCounter}
\stepcounter{AffiliationCounter}\edef\instMiyazaki{\protect\theAffiliationCounter}
\stepcounter{AffiliationCounter}\edef\instMEPhI{\protect\theAffiliationCounter}
\stepcounter{AffiliationCounter}\edef\instNagoya{\protect\theAffiliationCounter}
\stepcounter{AffiliationCounter}\edef\instNagoyaKMI{\protect\theAffiliationCounter}
\stepcounter{AffiliationCounter}\edef\instUNapoli{\protect\theAffiliationCounter}
\stepcounter{AffiliationCounter}\edef\instNara{\protect\theAffiliationCounter}
\stepcounter{AffiliationCounter}\edef\instNCU{\protect\theAffiliationCounter}
\stepcounter{AffiliationCounter}\edef\instNUU{\protect\theAffiliationCounter}
\stepcounter{AffiliationCounter}\edef\instTaiwan{\protect\theAffiliationCounter}
\stepcounter{AffiliationCounter}\edef\instKrakow{\protect\theAffiliationCounter}
\stepcounter{AffiliationCounter}\edef\instNihonDental{\protect\theAffiliationCounter}
\stepcounter{AffiliationCounter}\edef\instNiigata{\protect\theAffiliationCounter}
\stepcounter{AffiliationCounter}\edef\instNovaGorica{\protect\theAffiliationCounter}
\stepcounter{AffiliationCounter}\edef\instNovosibirsk{\protect\theAffiliationCounter}
\stepcounter{AffiliationCounter}\edef\instOsakaCity{\protect\theAffiliationCounter}
\stepcounter{AffiliationCounter}\edef\instOsaka{\protect\theAffiliationCounter}
\stepcounter{AffiliationCounter}\edef\instPNNL{\protect\theAffiliationCounter}
\stepcounter{AffiliationCounter}\edef\instPanjab{\protect\theAffiliationCounter}
\stepcounter{AffiliationCounter}\edef\instPeking{\protect\theAffiliationCounter}
\stepcounter{AffiliationCounter}\edef\instPittsburgh{\protect\theAffiliationCounter}
\stepcounter{AffiliationCounter}\edef\instPunjab{\protect\theAffiliationCounter}
\stepcounter{AffiliationCounter}\edef\instNPC{\protect\theAffiliationCounter}
\stepcounter{AffiliationCounter}\edef\instRIKENMSL{\protect\theAffiliationCounter}
\stepcounter{AffiliationCounter}\edef\instRIKEN{\protect\theAffiliationCounter}
\stepcounter{AffiliationCounter}\edef\instUSTC{\protect\theAffiliationCounter}
\stepcounter{AffiliationCounter}\edef\instSeoul{\protect\theAffiliationCounter}
\stepcounter{AffiliationCounter}\edef\instShoyaku{\protect\theAffiliationCounter}
\stepcounter{AffiliationCounter}\edef\instSoochow{\protect\theAffiliationCounter}
\stepcounter{AffiliationCounter}\edef\instSoongsil{\protect\theAffiliationCounter}
\stepcounter{AffiliationCounter}\edef\instSungkyunkwan{\protect\theAffiliationCounter}
\stepcounter{AffiliationCounter}\edef\instSydney{\protect\theAffiliationCounter}
\stepcounter{AffiliationCounter}\edef\instTabuk{\protect\theAffiliationCounter}
\stepcounter{AffiliationCounter}\edef\instTata{\protect\theAffiliationCounter}
\stepcounter{AffiliationCounter}\edef\instTUM{\protect\theAffiliationCounter}
\stepcounter{AffiliationCounter}\edef\instTelAviv{\protect\theAffiliationCounter}
\stepcounter{AffiliationCounter}\edef\instToho{\protect\theAffiliationCounter}
\stepcounter{AffiliationCounter}\edef\instTohoku{\protect\theAffiliationCounter}
\stepcounter{AffiliationCounter}\edef\instERI{\protect\theAffiliationCounter}
\stepcounter{AffiliationCounter}\edef\instTokyo{\protect\theAffiliationCounter}
\stepcounter{AffiliationCounter}\edef\instTIT{\protect\theAffiliationCounter}
\stepcounter{AffiliationCounter}\edef\instTMU{\protect\theAffiliationCounter}
\stepcounter{AffiliationCounter}\edef\instUtkal{\protect\theAffiliationCounter}
\stepcounter{AffiliationCounter}\edef\instVPI{\protect\theAffiliationCounter}
\stepcounter{AffiliationCounter}\edef\instWayneState{\protect\theAffiliationCounter}
\stepcounter{AffiliationCounter}\edef\instYamagata{\protect\theAffiliationCounter}
\stepcounter{AffiliationCounter}\edef\instYonsei{\protect\theAffiliationCounter}

\collaboration{The Belle Collaboration}
  \author[\instIITH]{S.~Choudhury,} % IITH
  \author[\instCincinnati,\instIITH]{S.~Sandilya,} % IITH
  \author[\instLAL]{K.~Trabelsi,} % LAL
  \author[\instIITH]{A.~Giri,} % IITH
  \author[\instTokyo]{H.~Aihara,} % Tokyo
  \author[\instTabuk,\instKAU]{S.~Al~Said,} % Tabuk
  \author[\instBNL]{D.~M.~Asner,} % BNL
  \author[\instCincinnati]{H.~Atmacan,} % Cincinnati
  \author[\instBINP,\instNovosibirsk]{V.~Aulchenko,} % BINP
  \author[\instHSE]{T.~Aushev,} % HSE
% ...

\title{\boldmath Test of lepton flavor universality and search for lepton flavor violation in \boldmath{$B \to K \ell\ell$} decays}

\preprint{\vbox{ \hbox{   }
					    	\hbox{Belle Preprint 2020-11}
                        	\hbox{KEK Preprint 2020-12} 
                     }}
                     
\abstract{
  We present measurements of the branching fractions for the decays $B\to K \mu^{+}\mu^{-}$ and $B\to K e^{+}e^{-}$, and their ratio ($R_{K}$), using a data sample of 711~\invfb  ~that contains $772 \times 10^{6}$ $B\bar{B}$ events. The data were collected at the \Y4S resonance with the Belle detector at the KEKB asymmetric-energy $e^{+}e^{-}$ collider. The ratio $R_{K}$ is measured in five bins of dilepton invariant-mass-squared ($q^{2}$): $q^{2} \in (0.1, 4.0), (4.00, 8.12), (1.0, 6.0)$, $(10.2, 12.8)$ and ($>14.18) \qq$, along with the whole $q^2$ region. The $R_{K}$ value for $q^{2} \in (1.0, 6.0) \qq$ is $1.03^{+0.28}_{-0.24} \pm 0.01$. The first and second uncertainties listed are statistical and systematic, respectively. All results for $R_{K}$ are consistent with Standard Model predictions. We also measure $C\!P$-averaged isospin asymmetries in the same $q^{2}$ bins. The results are consistent with a null asymmetry, with the largest difference of 2.6 standard deviations occurring for the $q^{2}\in(1.0,6.0)\qq$ bin in the mode with muon final states. The measured differential branching fractions, ${d\cal B}/{dq^{2}}$, are consistent with theoretical predictions for charged $B$ decays, while the corresponding values are below the expectations for neutral $B$ decays. We have also searched for lepton-flavor-violating $B \rightarrow K\mu^{\pm}e^{\mp}$ decays and set $90\%$ confidence-level upper limits on the branching fraction in the range of $10^{-8}$ for $B^{+} \rightarrow K^{+}\mu^{\pm}e^{\mp}$, and $B^{0} \rightarrow K^{0}\mu^{\pm}e^{\mp}$ modes.
}
 \usepackage{graphicx} 
 \usepackage{adjustbox}
\usepackage{pdflscape}
\usepackage{rotating}
\usepackage{longtable}
\usepackage{natbib}
\def\etal{{\it et al.}}
\def\invfb{\ensuremath{\mbox{\,fb}^{-1}}}
\def\Y#1S{\ensuremath{\Upsilon{(#1S)}}\;}
\newcommand{\gev}{\ensuremath{\mathrm{\,Ge\kern -0.1em V}}}
\newcommand{\mev}{\ensuremath{\mathrm{\,Me\kern -0.1em V}}\;}
\newcommand{\gevc}{\ensuremath{{\mathrm{\,Ge\kern -0.1em V\!/}c}}}
\newcommand{\mevc}{\ensuremath{{\mathrm{\,Me\kern -0.1em V\!/}c}}\;}
\newcommand{\gevcc}{\ensuremath{{\mathrm{\,Ge\kern -0.1em V\!/}c^2}}}
\newcommand{\qq}{\ensuremath{{\mathrm{\,Ge\kern -0.1em V^2\!/}c^4}}}
\newcommand{\mevcc}{\ensuremath{{\mathrm{\,Me\kern -0.1em V\!/}c^2}}}
\def\Mbc{\ensuremath{M^{}_{\rm bc}}}
\def\KS     {\ensuremath{K^0_{\scriptscriptstyle S}}}
\begin{document}
\maketitle
\flushbottom

\section{Introduction}
\label{sec:intro}

{\renewcommand{\thefootnote}{\fnsymbol{footnote}}}
\setcounter{footnote}{0}
The decays $B\to K\ell^{+}\ell^{-}$ ($\ell=e,\mu$), mediated by the $b\to s\ell^{+}\ell^{-}$ quark-level transition, constitute a flavor-changing neutral current process. Such processes are forbidden at tree level in the Standard Model (SM) but can proceed via suppressed loop-level diagrams, and they are therefore sensitive to particles predicted in a number of new physics models~\cite{{th:bsm:1},{th:bsm:2}}.
A robust observable~\cite{clean_obs} to test the SM prediction is the lepton-flavor-universality (LFU) ratio,
\begin{equation}
  R_{H} = \frac{\int\frac{d\Gamma}{dq^{2}}[ B\to H\mu^{+}\mu^{-}]dq^{2}}{\int\frac{d\Gamma}{dq^{2}}[B\to He^{+}e^{-}]dq^{2}},
\label{RK}
\end{equation}
where $H$ is a $K$ or $K^{\ast}$ meson and the decay rate $\Gamma$ is integrated over a range of the dilepton invariant mass squared, $q^{2}\equiv M^{2}(\ell^{+}\ell^{-})$. For $R_{K^{\ast}}$, recently LHCb~\cite{ex:lhcb:rkst} reported hints of deviations from SM expectations, while Belle~\cite{ex:belle:rkst} results are consistent with the SM with relatively larger uncertainties. LHCb also measured $R_{K}$~\cite{ex:lhcb:rk}, reporting a difference of about 2.5 standard deviations ($\sigma$) from the SM prediction in the $q^{2}\in(1.1,6.0)\qq$ bin. A previous measurement of the same quantity was performed by Belle~\cite{ex:belle:rk} in the whole $q^{2}$ range with a data sample of $657 \times 10^{6}$ $B\bar{B}$ events. The result presented here is obtained from a multidimensional fit performed on the full Belle data sample of $772 \times 10^{6}$ $B\bar{B}$ events, and supersedes our previous result~\cite{ex:belle:rk}.

Another theoretically robust observable~\cite{ai_theory}, where the dominant form-factor-related uncertainties also cancel, is the $C\!P$-averaged isospin asymmetry, representing the difference in partial widths: 
\begin{equation}
  A_{I} = \frac{({\tau_{B^{+}}}/{\tau_{B^{0}})}{\cal B}(B^{0}\to K^{0}\ell^{+}\ell^{-}) - {\cal B}(B^{+}\to K^{+}\ell^{+}\ell^{-})}{({\tau_{B^{+}}}/{\tau_{B^{0}})}{\cal B}(B^{0}\to K^{0}\ell^{+}\ell^{-}) + {\cal B}(B^{+}\to K^{+}\ell^{+}\ell^{-})},
\label{AI}
\end{equation}
where ${\tau_{B^{+}}}/{\tau_{B^{0}}} = 1.078\pm 0.004$ is the lifetime ratio of $B^{+}$ to $B^{0}$~\cite{pdg}. The $f^{+-}/f^{00} = 1.058 \pm 0.024$ \cite{ff_value} is the relative production fraction of charged ($f^{+-}$) and neutral $B$ mesons ($f^{00}$) at $B$ factories. The $A_{I}$ value is expected to be close to zero in the SM~\cite{th:ai}. Earlier, BaBar~\cite{ex:babar:ai} and Belle~\cite{ex:belle:rk} reported $A_{I}$ to be significantly below zero, especially in the $q^{2}$ region below the $J/\psi$ resonance, while LHCb~\cite{ex:lhcb:ai} reported results consistent with SM predictions.

In many theoretical models, lepton flavor violation (LFV) accompanies LFU violation~\cite{lfv_lfuv}. With neutrino mixing, LFV is only possible at rates far below the current experimental sensitivity. In case of signal, this will signify physics beyond SM~\cite{neutrino_lfv}. The LFV in $B$ decays can be studied via $B \rightarrow K\mu^{\pm}e^{\mp}$. The most stringent upper limits on $B^{+} \rightarrow K^{+}\mu^{+}e^{-}$ and $B^{+} \rightarrow K^{+}\mu^{-}e^{+}$ set by LHCb~\cite{LHCb_lfv} are $6.4 \times 10^{-9}$ and $7.0 \times 10^{-9}$ at $90\%$ confidence level (CL). Prior to that, $B^{0} \rightarrow K^{0}\mu^{\pm}e^{\mp}$ decays were searched for by BaBar~\cite{BaBar_lfv}, which set a 90\% CL upper limit on the branching fraction of $2.7 \times 10^{-7}$. 

In this paper, we report a measurement of the decay branching fractions of $B\to K\ell^{+}\ell^{-}$, $R_{K}$ and $A_{I}$ in the whole $q^{2}$ range as well as in five $q^{2}$ bins [(0.1, 4.0), (4.00, 8.12), (1.0, 6.0), (10.2, 12.8) and ($>14.18$)] \qq. We also search for $B \to K \mu^{\pm} e^{\mp}$ decays using the full Belle data sample.

\section{Data samples and Belle detector}
\label{Sec:datadetector}

 This analysis uses a 711 \invfb~data sample containing $(772\pm 11) \times 10^6$ $B\bar{B}$ events, collected at the $\Upsilon(4S)$ resonance by the Belle experiment at the KEKB $e^{+}e^{-}$ collider~\cite{KEKB}. An 89\invfb~data sample recorded 60\mev below the \Y4S peak (off-resonance) is used to estimate the background contribution from $e^{+}e^{-}\to q\bar{q}$ ($q = u,d,s, c$) continuum events.

The Belle detector~\cite{belle:detector} is a large-solid-angle magnetic spectrometer composed of a silicon vertex detector (SVD), a 50-layer central drift chamber (CDC), an array of aerogel threshold Cherenkov counters (ACC), a barrel-like arrangement of time-of-flight scintillation counters (TOF), and an electromagnetic calorimeter (ECL) comprising CsI(Tl) crystals. All these subdetectors are located inside a superconducting solenoid coil that provides a 1.5~T magnetic field. An iron flux-return yoke placed outside the coil is instrumented with resistive plate chambers (KLM) to detect $K^{0}_{L}$ mesons and muons. Two inner detector configurations were used: a 2.0 cm radius beam pipe and a three-layer SVD for the first sample of 140\invfb; and a 1.5 cm radius beam pipe, a four-layer SVD, and a small-cell inner CDC for the remaining 571\invfb~\cite{nbb}.

To study properties of signal events and optimize selection criteria, we use samples of Monte Carlo (MC) simulated events. The $B \to K \ell^+\ell^-$ modes are generated with the \textsc{EvtGen} package~\cite{evtgen} based on a model described in Ref.~\cite{btosllball}, while LFV modes are generated with a phase-space model. The \textsc{PHOTOS}~\cite{photos} package is used to incorporate final-state radiation. The detector response is simulated with \textsc{GEANT3}~\cite{geant3}. 

\section{Analysis Overview}
\label{Sec:ana}
We reconstruct $B\to K \ell^{+} \ell^{-}$ ($K=K^{+},\KS$)~\cite{cc} decays by selecting charged particles that originate from the vicinity of the $e^+e^-$ interaction point (IP), except for those originating from $\KS$ decays. We require impact parameters less than $1.0$~cm in the transverse plane and less than $4.0$~cm along the $z$ axis (parallel to the $e^{+}$ beam). To reduce backgrounds from low-momentum particles, we require that tracks have a minimum transverse momentum of 100\mevc.

From the list of selected tracks, we identify $K^{+}$ candidates using a likelihood ratio ${\cal R}^{}_{K/\pi} = {\cal L}^{}_K / ({\cal L}^{}_K + {\cal L}^{}_\pi )$, where ${\cal L}^{}_{K}$ and ${\cal L}^{}_{\pi}$ are the likelihoods for charged kaons and pions, respectively, calculated based on the number of photoelectrons in the ACC, the specific ionization in the CDC, and the time of flight as determined from the TOF. 
We select kaons by requiring ${\cal R}^{}_{K/\pi} > 0.6$, which has a kaon identification efficiency of 92\% and a pion misidentification rate of 7\%. 
For the neutral $B$ decay, candidate \KS~mesons are reconstructed by combining two oppositely charged tracks (assumed to be pions) with an invariant mass between $487$ and $508$ $\mevcc$; this corresponds to a $3\sigma$ window around the  nominal \KS~mass~\cite{pdg}. Such candidates are further identified with a neural network (NN). 
The variables used for this NN are: the \KS~momentum; the distance along the $z$ axis between the two track helices at their closest approach; the flight length in the transverse plane; the angle between the \KS~momentum and the vector joining the IP with the \KS~decay vertex; the angle between the pion momentum and the laboratory-frame direction in the \KS~rest frame; the distances-of-closest-approach in the transverse plane between the IP and the two pion helices; and the number of hits in the CDC; and the presence or absence of hits in the SVD for each pion track.   

Muon candidates are identified based on information from the KLM. We require that candidates have a momentum greater than 0.8~\gevc~(enabling them to reach the KLM subdetector), and a penetration depth and degree of transverse scattering consistent with those of a muon~\cite{muid}. The latter information is used to calculate a normalized muon likelihood ${\cal R}_{\mu}$, and we require ${\cal R}_{\mu} > 0.9$. For this requirement, the average muon detection efficiency is 89\%, with a pion misidentification rate of 1.5\%~\cite{pid}.

Electron candidates are required to have a momentum greater than 0.5~\gevc~and are identified using the ratio of calorimetric cluster energy to the CDC track momentum; the shower shape in the ECL; the matching of the track with the ECL cluster; the specific ionization in the CDC; and the number of photoelectrons in the ACC.
This information is used to calculate a normalized electron likelihood ${\cal R}^{}_{e}$, and we require ${\cal R}^{}_{e} >~0.9$.
This requirement has an efficiency of 92\% and a pion misidentification rate below 1\%~\cite{eid}. 
To recover energy loss due to possible bremsstrahlung, we search for photons inside a cone of radius $50~\rm mrad$ centered around the electron direction. For each photon found within the cone, its four-momentum is added to that of the initial electron.

Charged (neutral) $B$~candidates are reconstructed by combining $K^{\pm}$ (\KS) with suitable $\mu^\pm$ or $e^\pm$ candidates. To distinguish signal from background events, two kinematic variables are used: the beam-energy-constrained mass $\Mbc = \sqrt{(E^{}_{\rm beam}/c^{2})^{2} - (p^{}_{B}/c)^{2}}$, and the energy difference $\Delta E =  E^{}_{B} - E^{}_{\rm beam}$, where $E^{}_{\rm beam}$ is the beam energy, and  $E^{}_{B}$ and $p^{}_{B}$ are the energy and momentum, respectively, of the $B$ candidate. All these quantities are calculated in the $e^{+}e^{-}$ center-of-mass (CM) frame. For signal events, the $\Delta E$ distribution peaks at zero, and the \Mbc~distribution peaks near the $B$ mass. We retain candidates satisfying the requirements $-0.10 < \Delta E < 0.25 \gev$ and $\Mbc > 5.2~\gevcc$.

With the above selection criteria applied, about 2\% of signal MC events are found to have more than one $B$ candidate. 
For these events, we retain the candidate with smallest $\chi^{2}$ value resulting from a vertex fit of the $B$ daughter particles. 
From MC simulation, this criterion is found to select the correct signal candidate 78-85\% of the time, depending on the decay mode. The decays $B\to J/\psi (\to \ell^{+}\ell^{-}) K$ and $B\to \psi (2S) (\to \ell^{+}\ell^{-})K$, used later as control samples, are suppressed in the signal selection via a set of vetoes $8.75 < q^{2} < 10.2\qq$ and $13.0 < q^{2} < 14.0\qq$ with the dimuon; $8.12 < q^{2}<10.2\qq$ and $12.8<q^{2}<14.0\qq$ with the dielectron final states for $B\to  J/\psi K$ and $B\to \psi (2S) K$, respectively. An additional veto of the low $q^{2}$ region ($< 0.05$\qq) is applied in the case of $B\to K e^{+}e^{-}$ to suppress possible contaminations from $\gamma^{\ast}\to e^{+} e^{-}$ and $\pi^{0}\to\gamma e^{+} e^{-}$. 

At this stage of the analysis, there is significant background from  continuum processes and other $B$ decays. 
As lighter quarks are produced with large kinetic energy, the former events tend to consist of two back-to-back jets of pions and kaons. 
In contrast, $B\bar{B}$ events are produced almost at rest in the CM frame, resulting in more spherically distributed daughter particles. 
We thus distinguish $B\bar{B}$ events from $q\bar{q}$ background based on event topology.

Background arising from $B$ decays has typically two uncorrelated leptons in the final state. Such background falls into three categories: ({\it a})\ both $B$ and $\bar{B}$ decay semileptonically; ({\it b})\ a $B\to \bar{D}^{(*)}X\ell^+\nu$ decay is followed by $\bar{D}^{(*)}\to X\ell^-\bar{\nu}$; and ({\it c})\ hadronic $B$ decays where one or more daughter particles are misidentified as leptons.  To suppress continuum as well as $B\bar{B}$ background, we use an NN trained with the following input variables:

\begin{enumerate}

\item A likelihood ratio constructed from a set of modified Fox-Wolfram moments~\cite{{KSFW},{FW}}.

\item The angle $\theta^{}_B$ between the $B$ flight direction and the $z$ axis in the CM frame (for $B\bar{B}$ events, $dN/d\cos\theta^{}_B\propto 1-\cos^{2}\theta_{B}$, whereas for continuum events, $dN/d\cos\theta^{}_B\approx {\rm\ constant}$, where $N$ is the number of events).

\item The angle $\theta^{}_T$ between the thrust axes calculated from final-state particles for the candidate $B$ and for the rest of the event in the CM frame. (The thrust axis is the direction that maximizes the sum of the longitudinal momenta of the considered particles). For signal events, the  $\cos\theta^{}_T$ distribution is flat, whereas for continuum events it peaks near $\pm 1$.

\item Flavor-tagging information from the tag-side (recoiling) $B$ decay. The flavor-tagging algorithm~\cite{belle:qr} outputs two variables: the flavor $q$ of the tag-side $B$, and the tag quality~$r$. The latter ranges from zero for no flavor information to one for an unambiguous flavor assignment.

\item The confidence level of the $B$ vertex fitted from all daughter particles.

\item The separation in $z$ between the signal $B$ decay vertex and that of the other $B$ in the event.

\item The separation between the two leptons along the $z$-axis divided by the quadratic sum of uncertainties in the $z$-intercepts of the tracks.

\item The sum of the ECL energy of tracks and clusters not associated with the signal $B$ decay.
  
\item A set of variables developed by CLEO~\cite{cleocones} that characterize the momentum flow into concentric areas around the thrust axis of a reconstructed $B$ candidate.
\end{enumerate}

The NN outputs a single variable $\cal O$, for which larger values correspond to more signal-like events. To facilitate modeling of the distribution of $\cal O$ with an analytic function, we require ${\cal O} > -0.6~(={\cal O}_{\rm min})$ and transform $\cal O$ to a new variable:
\begin{equation*}
  {\cal O'} = \log\left[\frac{{\cal O} - {\cal O}_{\rm min}}{{\cal O}_{\rm max} - {\cal O}}\right],
\end{equation*}
where ${\cal O}_{\rm max}$ is the upper boundary of $\cal O$. The criterion on  $\cal O_{\rm min}$ reduces the background events by more than 75\%, with a signal loss of only 4-5\%. 

We study the remaining background events using MC simulation for individual modes, with special attention paid to those that can mimic signal decays. 
Candidates arising from $B^{0}\to J/\psi (\rightarrow \ell^{+}\ell^{-}) K^{\ast 0}$ populate towards the negative side in $\Delta E$ and are suppressed with the requirement $\Delta E > -0.1\gev$. 
The decay $B^{+}\to \bar{D}^{0}(\rightarrow K^{+}\pi^{-})\pi^{+}$ mimics $B^{+}\to K^{+}\mu^{+}\mu^{-}$ when both pions are mis-identified as muons; to suppress this background, we apply a veto on the invariant mass of the $K^{+}$ and $\mu^{-}$ candidates: $M[K^{+}\mu^{-}] \notin (1.85,1.88)\gevcc$. 
The contribution from other $B$ $\to$ charm decays is negligible. 
Events originating from the decays $B^{+}\to J/\psi (\rightarrow \mu^{+}\mu^{-}) K^{+}$, in which one of the muons is misidentified as a kaon and vice versa, contribute as a peaking background to $B^{+}\to  K^{+}\mu^{+}\mu^{-}$. 
Such events are suppressed by applying a veto on the invariant mass $M[K^{+}\mu^{-}]\notin (3.06,3.13)\gevcc$. 

For the LFV modes, the background coming from $B^{+} \rightarrow J/\psi(\rightarrow e^+e^-) K^{+}$ because of misidentification and swapping between particles is removed by invariant mass vetoes. For the $B^{+} \rightarrow K^{+}\mu^{+}e^{-}$ mode, two vetoes are applied: (a) the electron is misidentified as kaon and kaon as muon, and thus the veto on the kaon-electron invariant mass is $M[K^{+}e^{-}] \notin (2.95, 3.11)\gevcc$; and (b) the electron is misidentified as a muon, and thus the muon-electron mass veto is $M[\mu^{+}e^{-}] \notin (3.02, 3.12)\gevcc$. For the $B^{+} \rightarrow K^{+}\mu^{-}e^{+}$ channel, only the latter background is found and removed using $M[\mu^{+}e^{-}] \notin (3.02, 3.12)\gevcc$. A small contribution from $B^{+}\to \bar{D}^{0}(K^{+}\pi^{-})\pi^{+}$ for these LFV modes, due to misidentification of pions as leptons, is removed by requiring $M[K^{+}\mu^{-}] \notin (1.85,1.88)\gevcc$. For the $B^{0} \rightarrow \KS \mu^{+}e^{-}$ mode, a background contribution from $B^{0}\rightarrow J/\psi(\rightarrow e^+e^-) \KS$, where an electron is misrecontructed as a muon, is suppressed by requiring $M[\mu^{+}e^{-}] \notin (3.04, 3.12)\gevcc$. When calculating invariant masses for these vetoes, the mass hypothesis for the misidentified particle is used. 
There is a small background from $B \rightarrow K\pi^{+}\pi^{-}$ decays in the $B^{+}\rightarrow K^{+}\mu^{+}\mu^{-}$ ($1.37 \pm 0.03$ events), $B^{0}\rightarrow \KS\mu^{+}\mu^{-}$ ($0.17 \pm 0.01$ events), $B^{+}\rightarrow K^{+}\mu^{+}e^{-}$ ($0.16\pm 0.03$ events), $B^{+}\rightarrow K^{+}\mu^{-}e^{+}$ ($0.14\pm 0.03$ events), and $B^{0}\rightarrow \KS\mu^{+}e^{-}$ ($0.14\pm 0.02$ events) samples. This background is negligible in the $B^{+} \rightarrow K^{+}e^{+}e^{-}$ and $B^{0}\rightarrow \KS e^{+}e^{-}$ samples. 
The mentioned yields of peaking charmless $B$ backgrounds are estimated by considering all known intermediate resonances. To avoid biasing our results, all selection criteria are determined in a ``blind'' manner, {\it i.e.,} they are finalized before looking at data events in the signal region.

We determine the signal yields by performing a three-dimensional unbinned extended maximum-likelihood fit to the \Mbc, $\Delta E$, and ${\cal O'}$ distributions in different $q^{2}$ bins. 
The fits are performed for each mode separately. 
The probability density functions (PDFs) used to model signal decays are as follows: for \Mbc~we use a Gaussian, for $\Delta E$ the sum of a Gaussian and a Crystal Ball function~\cite{crystalball}, and for $\cal O'$ the sum of a Gaussian and an asymmetric Gaussian with a common mean. 
All signal shape parameters are obtained from MC simulation. 
To account for small differences observed between data and MC simulations, we introduce an offset in the mean positions and scaling factors for the widths. 
The values of these parameters are obtained from fitting the control sample $B\to J/\psi (\to\!\ell^{+}\ell^{-}) K$ decays and kept fixed. The PDFs used for charmless $B \rightarrow K\pi^{+}\pi^{-}$ peaking background is the same as that of the signal PDFs, with the fixed number of peaking events.
The shapes of the \Mbc, $\Delta E$, and ${\cal O'}$ distributions for background arising from $B$ decays are parameterized with an ARGUS function~\cite{argus}, an exponential, and a Gaussian function, respectively. 
Similarly, the continuum background is modeled using an ARGUS, a first-order polynomial, and a Gaussian function. 
The shapes of $B\bar{B}$ and continuum backgrounds are very similar in two of the fit variables, making it difficult to simultaneously float the yields of both backgrounds. 
Hence, the continuum yields are obtained for each mode from the off-resonance data sample and fixed in the fit. 
These yields are consistent with those of the high-statistics off-resonance MC sample. The $B\bar{B}$ yields are floated in the fit.

\section{Results}
\begin{figure}[hhh]
  \mbox{
    \includegraphics[width=0.33\columnwidth]{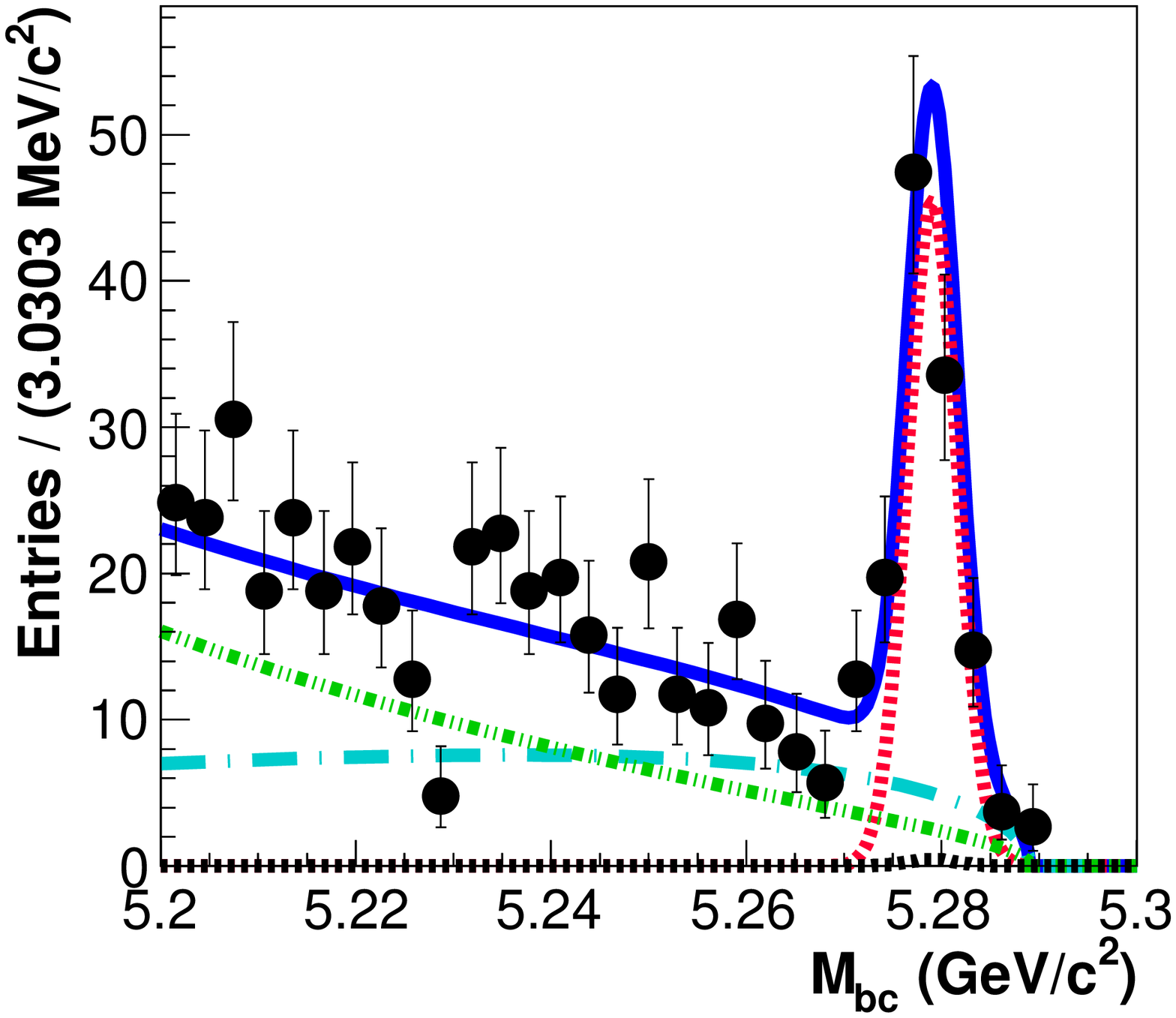}
    \includegraphics[width=0.33\columnwidth]{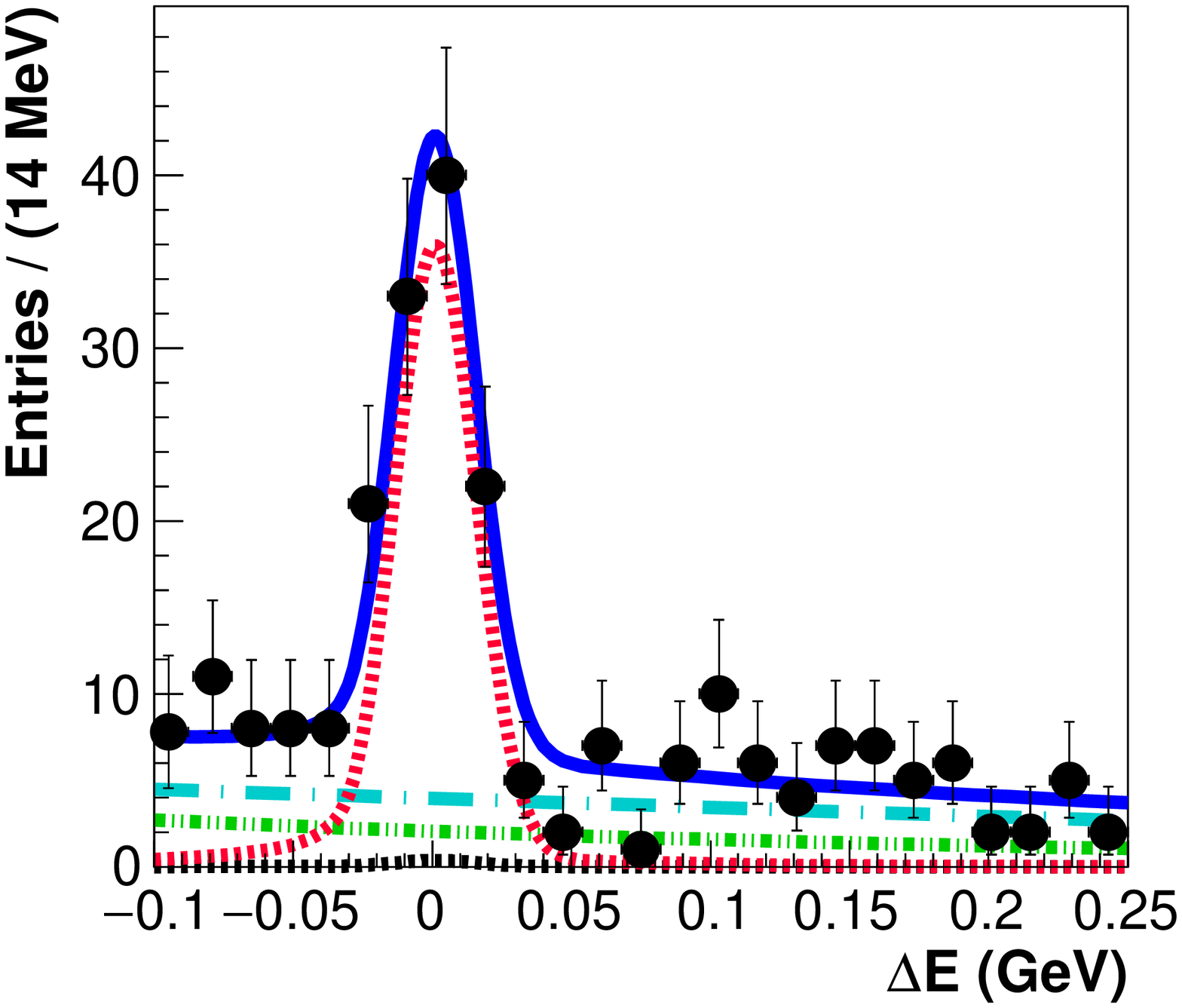}
    \includegraphics[width=0.33\columnwidth]{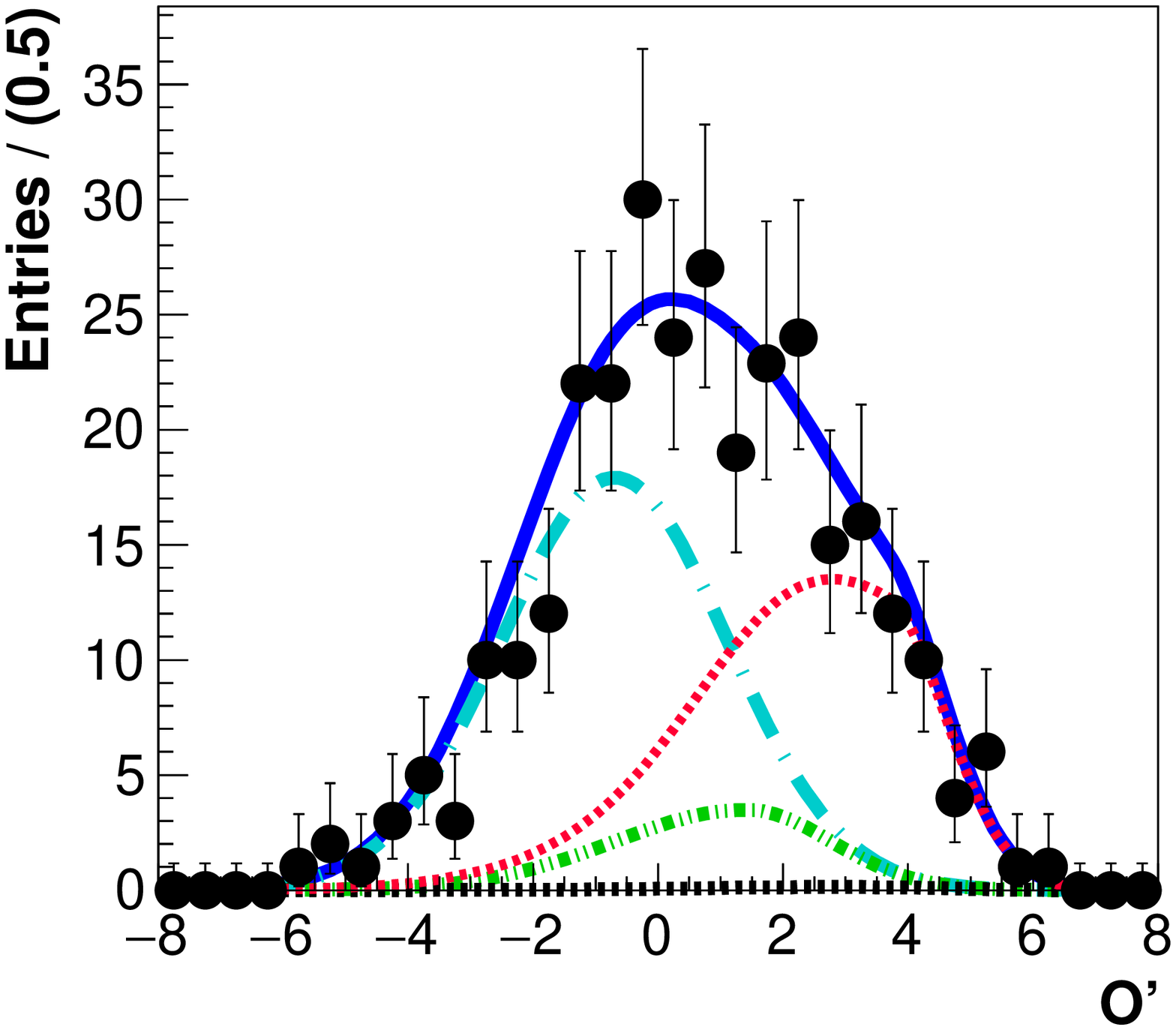}
    }
    
\mbox{
    \includegraphics[width=0.33\columnwidth]{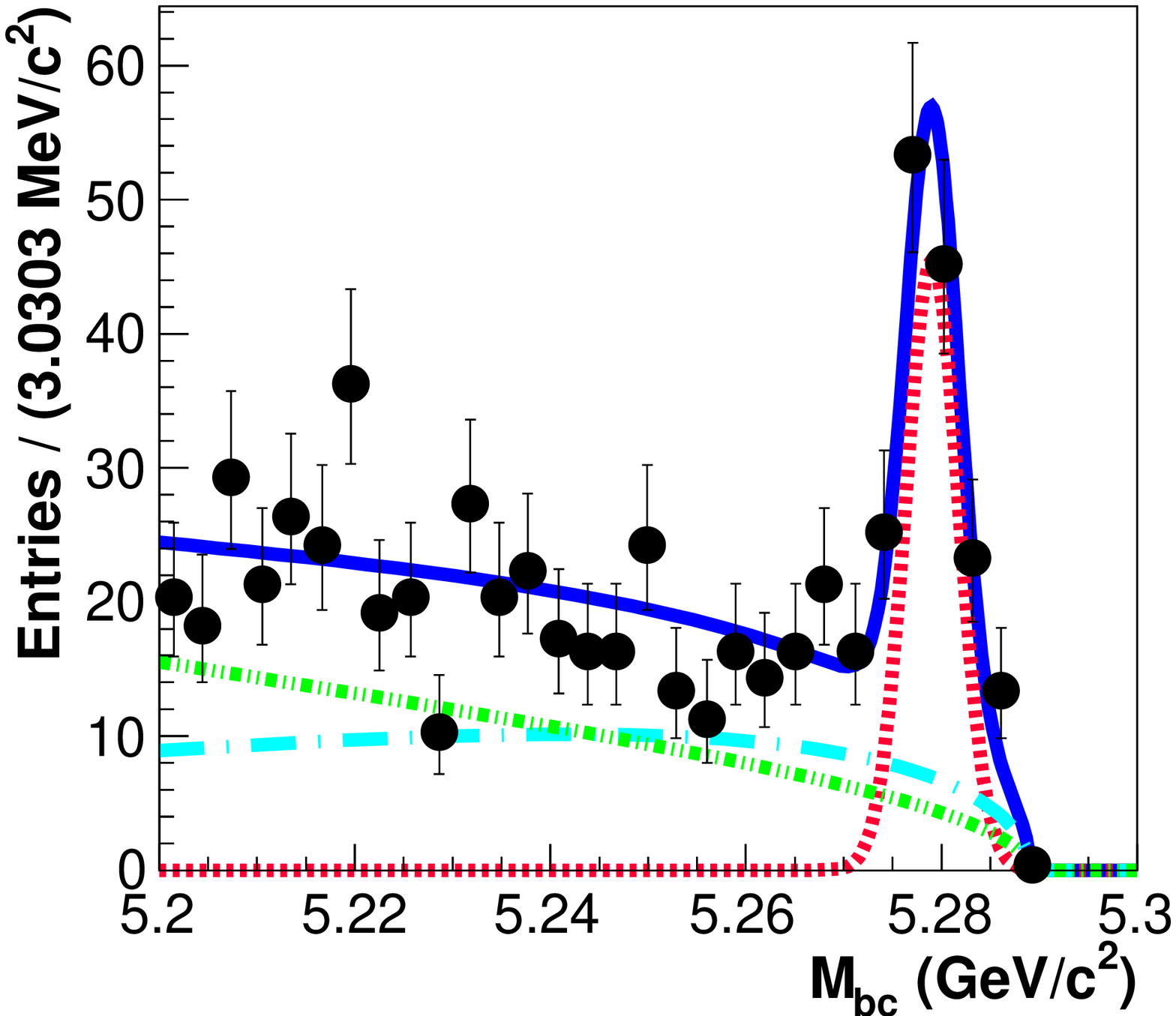}
    \includegraphics[width=0.33\columnwidth]{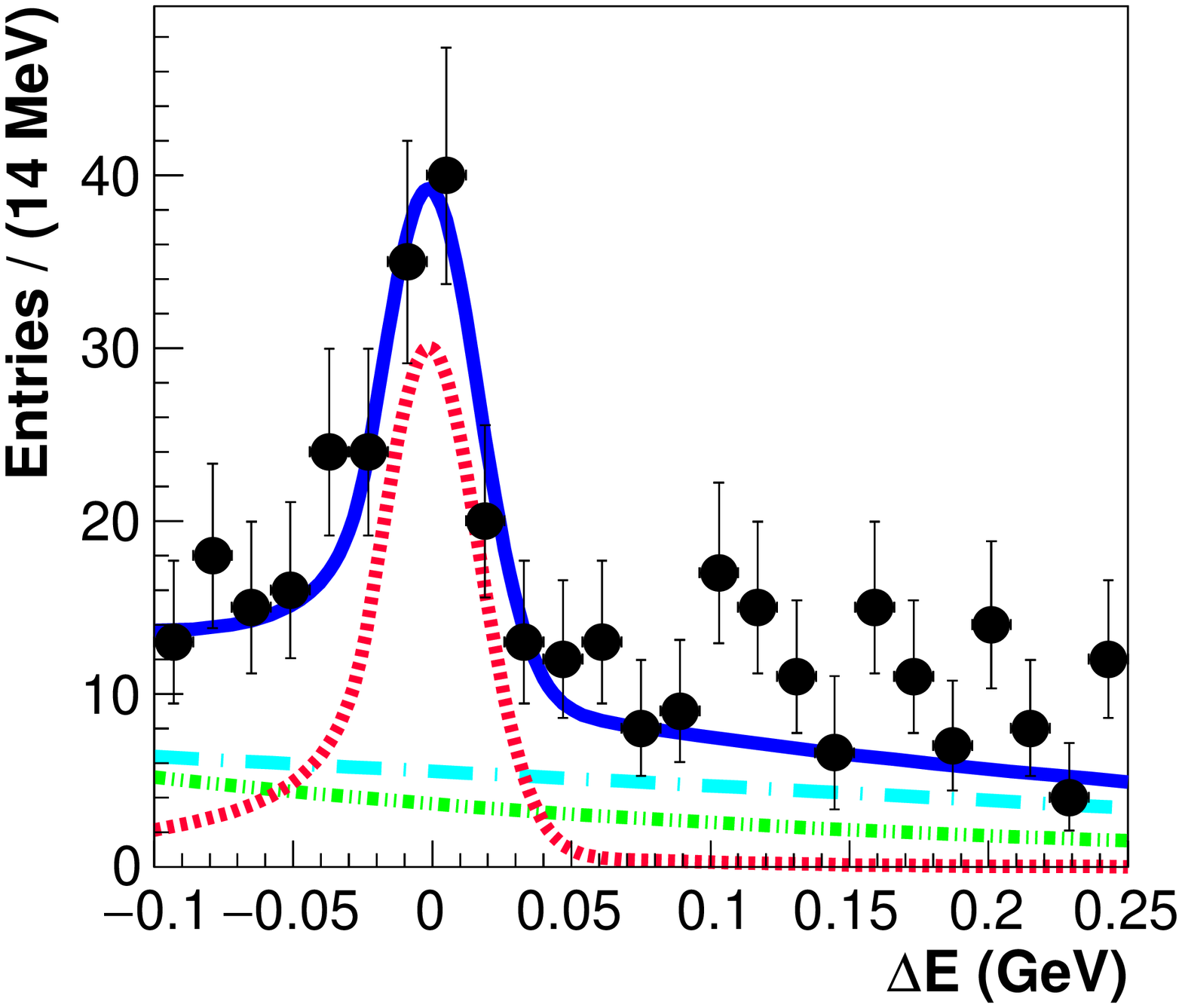}
    \includegraphics[width=0.33\columnwidth]{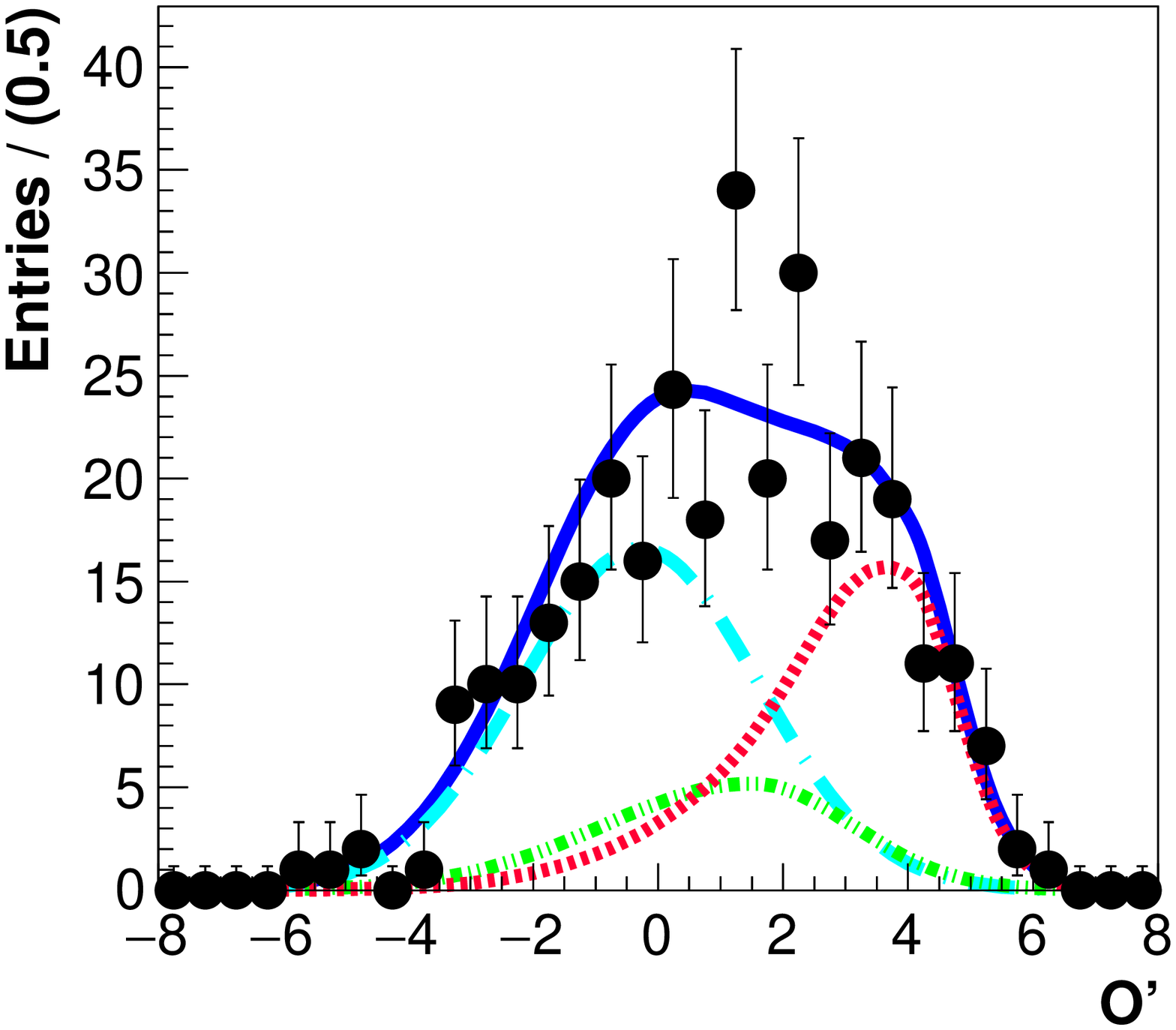}
    }
  \caption{Signal-enhanced \Mbc~(left), $\Delta E$ (middle), and ${\cal O'}$ (right) projections of three-dimensional unbinned extended maximum-likelihood fits to the data events that pass the selection criteria for $B^{+} \to K^{+} \mu^{+} \mu^{-}$ (top), and $B^{+} \to K^{+} e^{+} e^{-}$ (bottom). Points with error bars are the data; blue solid curves are the fitted results for the signal-plus-background hypothesis; red dashed curves denote the signal component; cyan long dashed, green dash-dotted, and black dashed curves represent continuum, $B\bar{B}$ background, and $B \rightarrow \rm{charmless}$ decays, respectively.}
  \label{fig:btokpll}
\end{figure}

\begin{figure}[htbp]
  \mbox{
    \includegraphics[width=0.33\columnwidth]{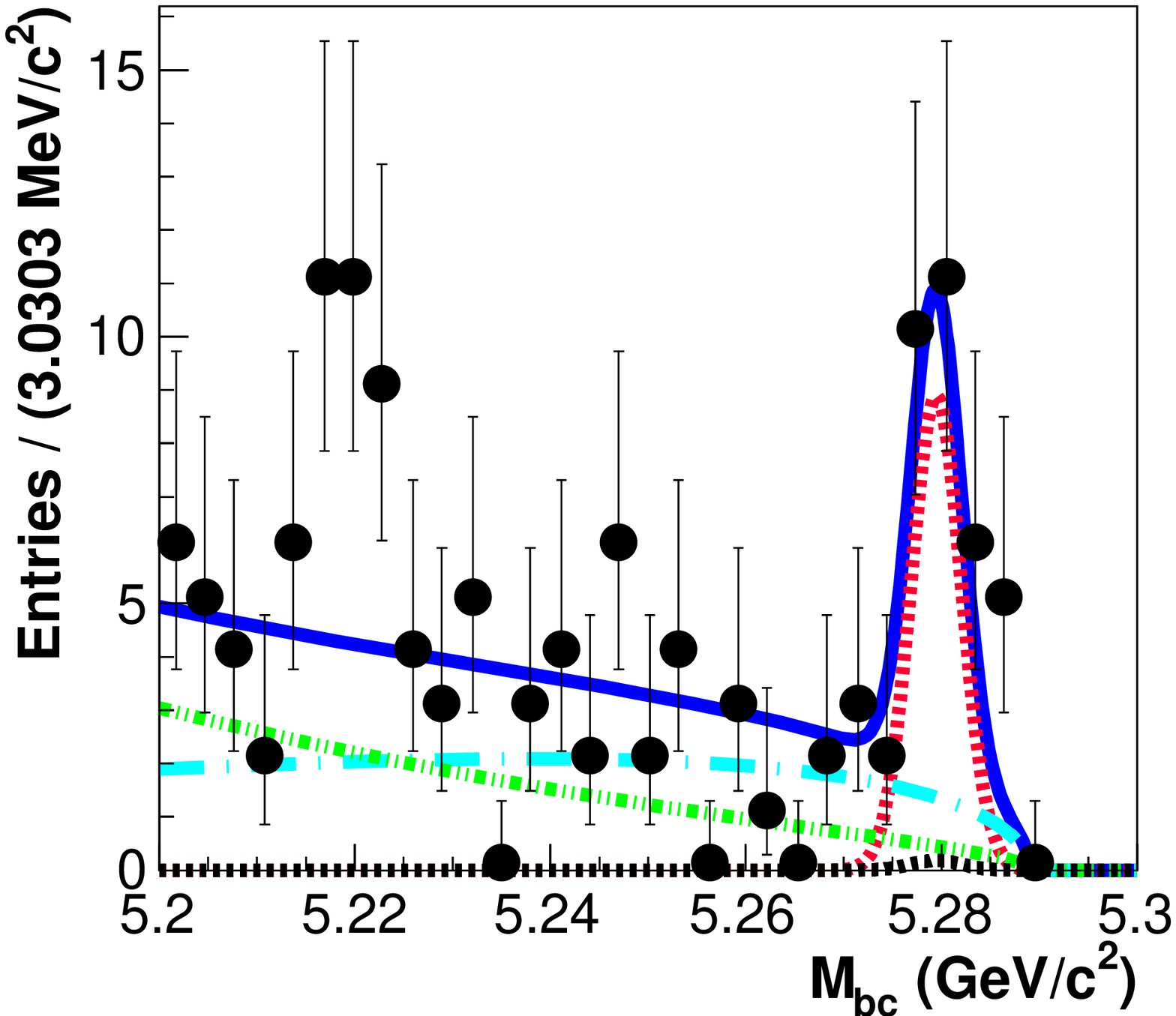}
    \includegraphics[width=0.33\columnwidth]{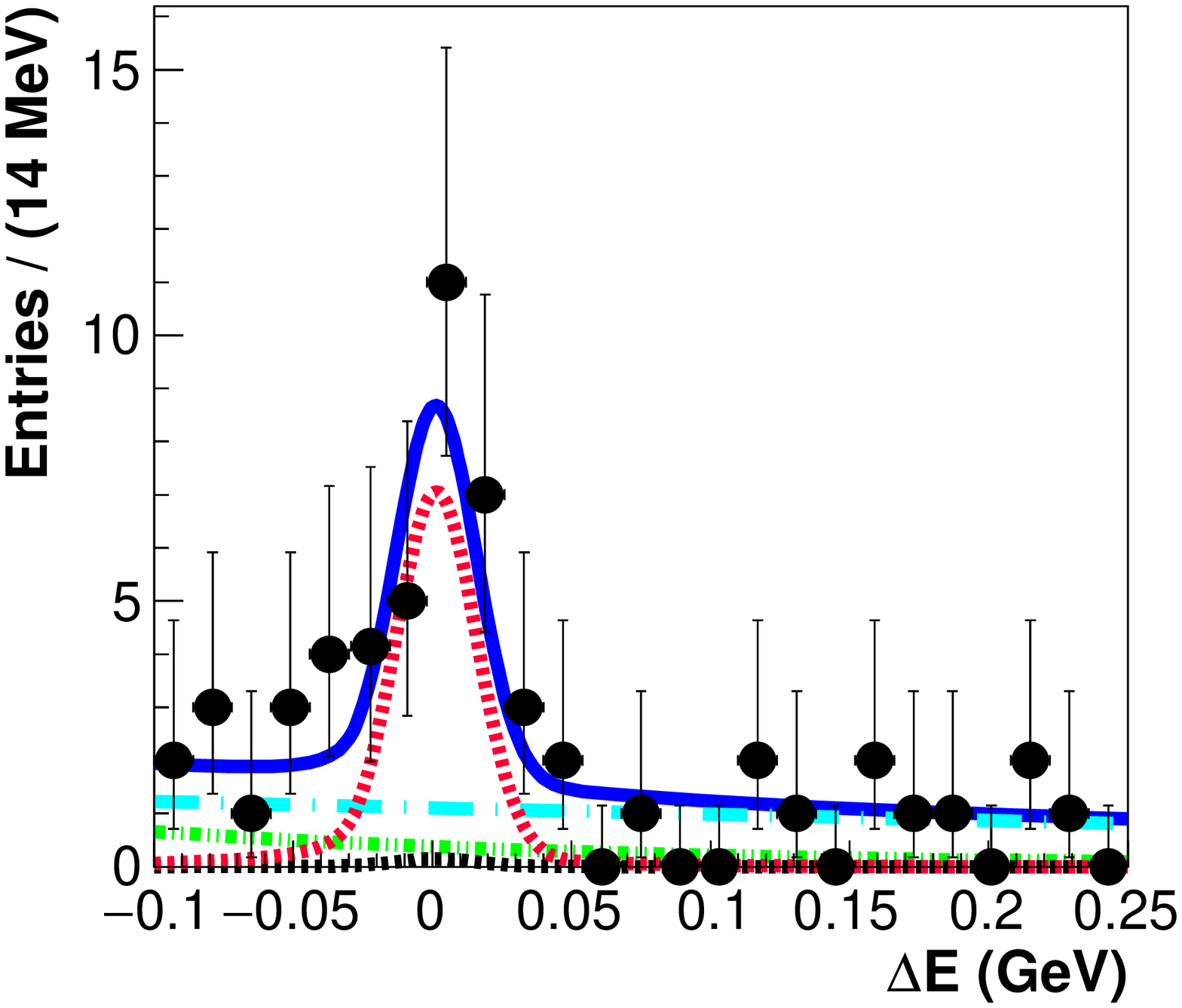}
    \includegraphics[width=0.33\columnwidth]{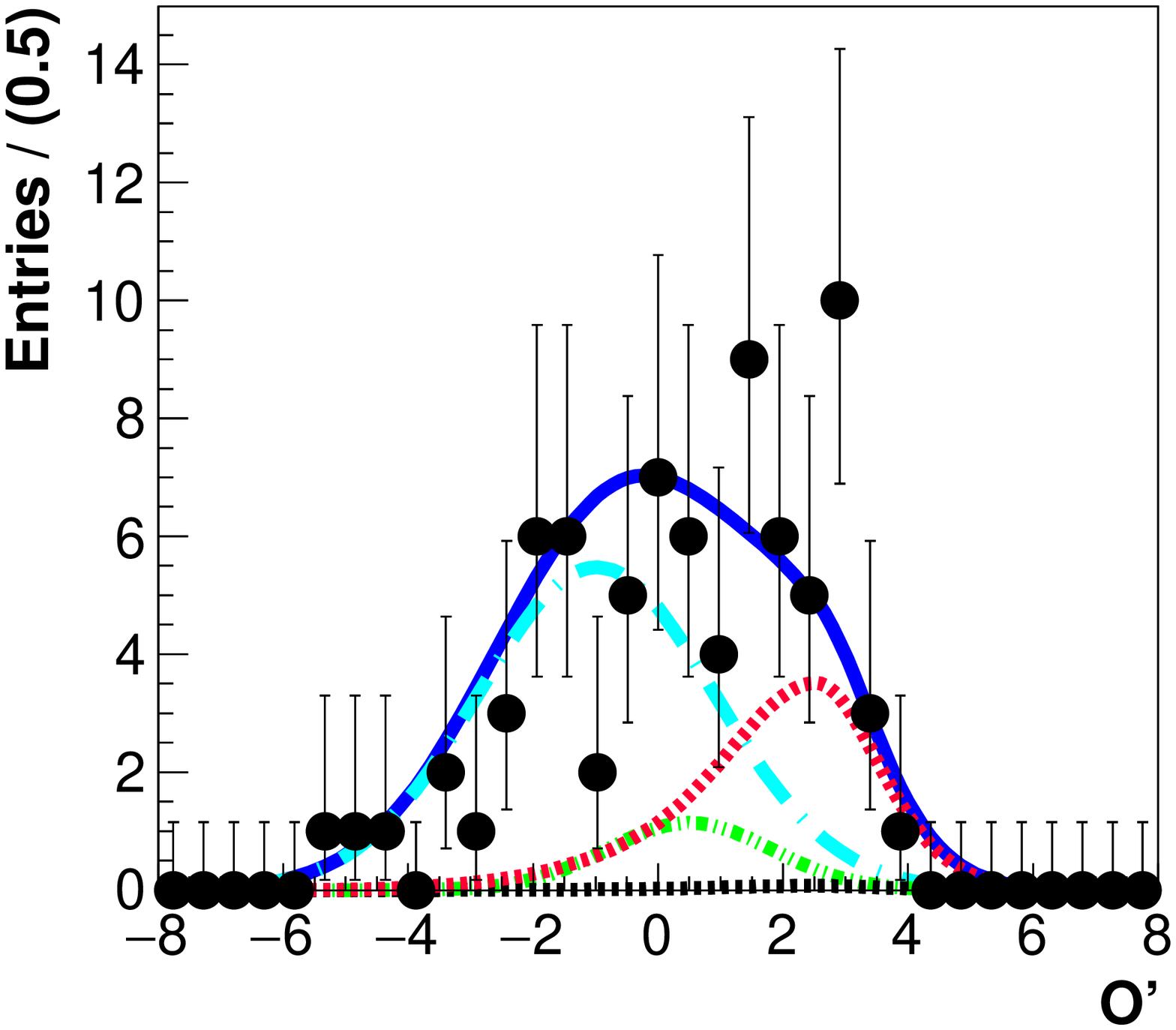}
    }
\mbox{
    \includegraphics[width=0.33\columnwidth]{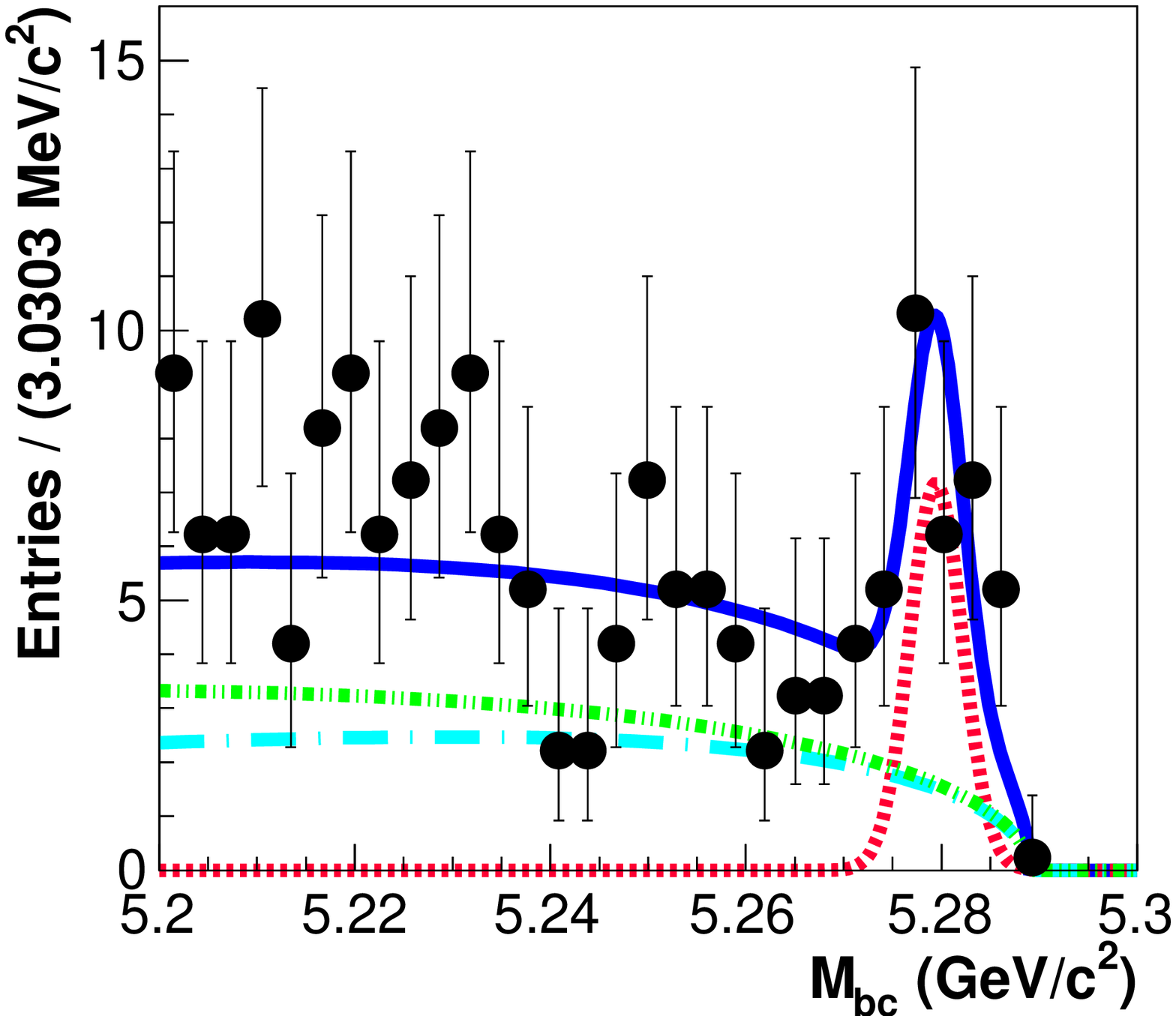}
    \includegraphics[width=0.33\columnwidth]{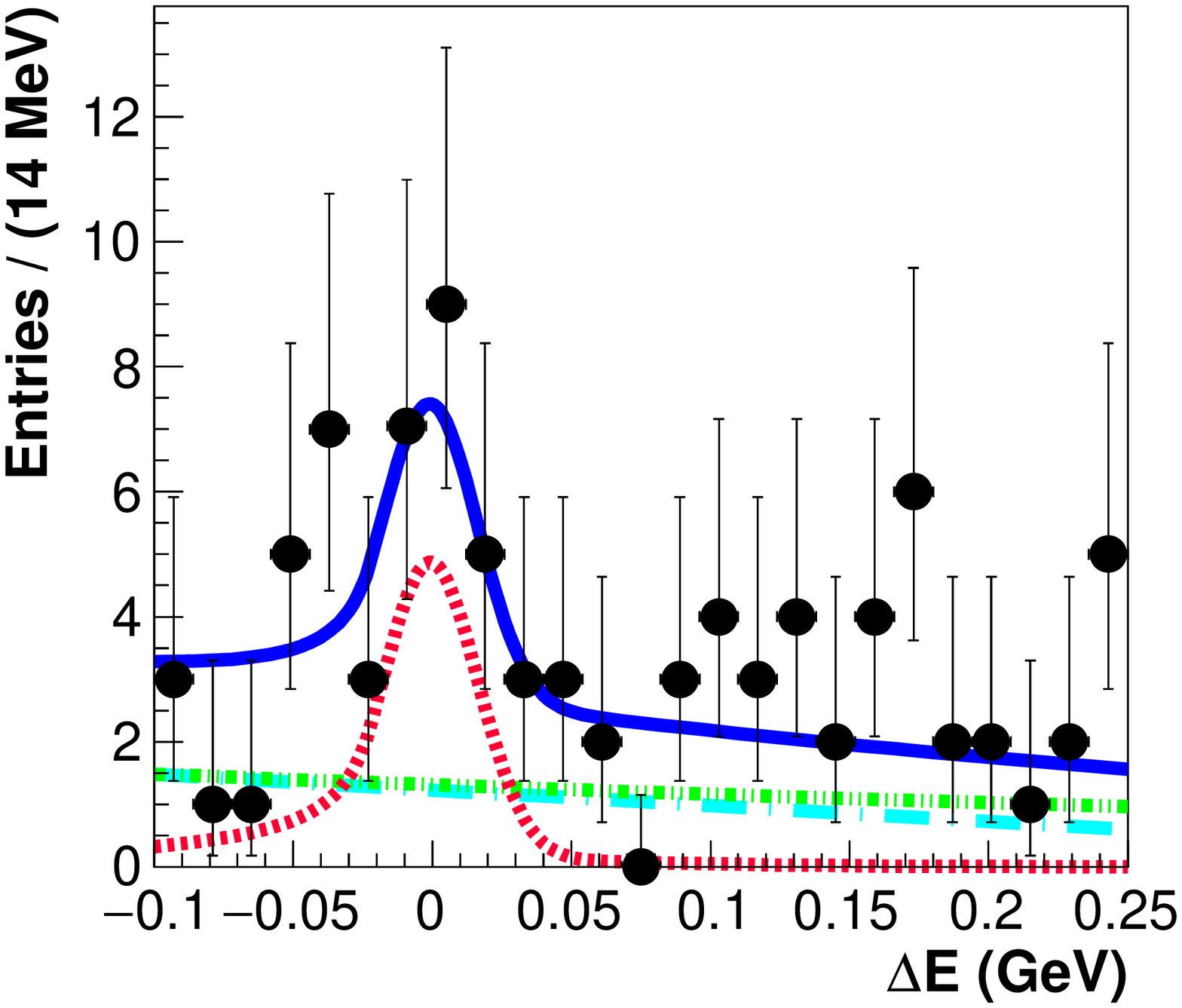}
    \includegraphics[width=0.33\columnwidth]{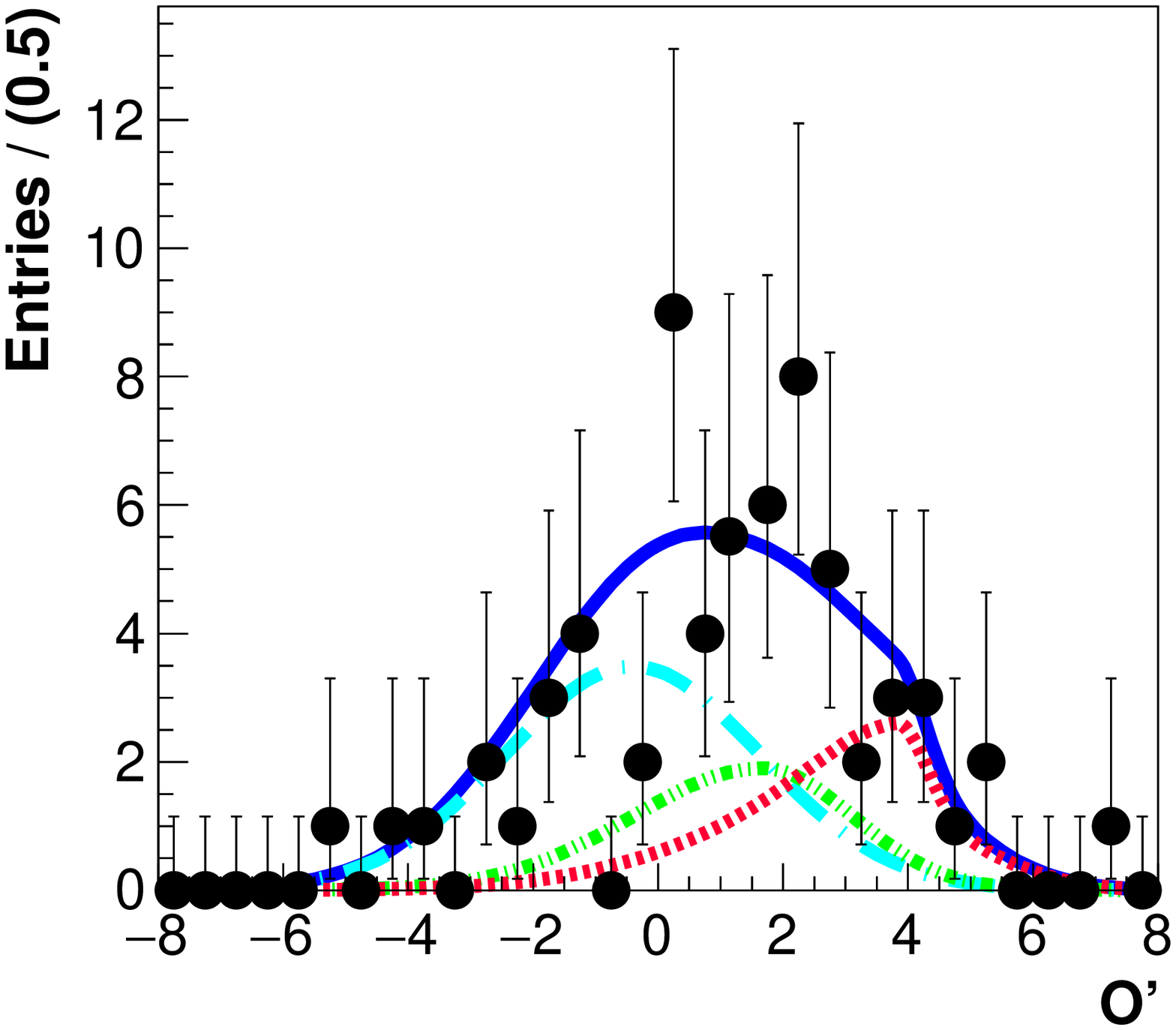}
    }
  \caption{Signal-enhanced \Mbc~(left), $\Delta E$ (middle), and ${\cal O'}$ (right) projections of three-dimensional unbinned extended maximum-likelihood fits to the data events that pass the selection criteria for $B^{0} \to \KS \mu^{+} \mu^{-}$ (top), and $B^{0} \to \KS e^{+} e^{-}$ (bottom). The legends are the same as in Fig. \ref{fig:btokpll}.}
  \label{fig:btoksll}
\end{figure}

The results of the fit projected into a signal-enhanced region for \Mbc~[$|\Delta E| < 0.05 \gev$ and ${\cal O'} \in (1.0,8.0)$], $\Delta E$ $[\Mbc \in (5.27,5.29) \gevcc$ and ${\cal O'} \in (1.0,8.0)]$ and ${\cal O'}$ $[\Mbc \in (5.27,5.29) \gevcc$ and $|\Delta E| < 0.05 \gev$] distributions in the data sample are shown in Figs.~\ref{fig:btokpll} and \ref{fig:btoksll} for $B^+ \rightarrow K^+ \ell^{+}\ell^{-}$ and $B^{0} \rightarrow \KS\ell^{+}\ell^{-}$, respectively. These distributions correspond to the whole $q^{2}$; $q^2 \in [(0.1, 8.75) (10.2, 13) (>14.18)$] \qq~with muon and $q^2 \in [(0.1, 8.12) (10.2, 12.8) (>14.18)$] \qq~with electron, in the final states. 


\begin{figure}[htbp]
  \mbox{
    \includegraphics[width=0.5\columnwidth]{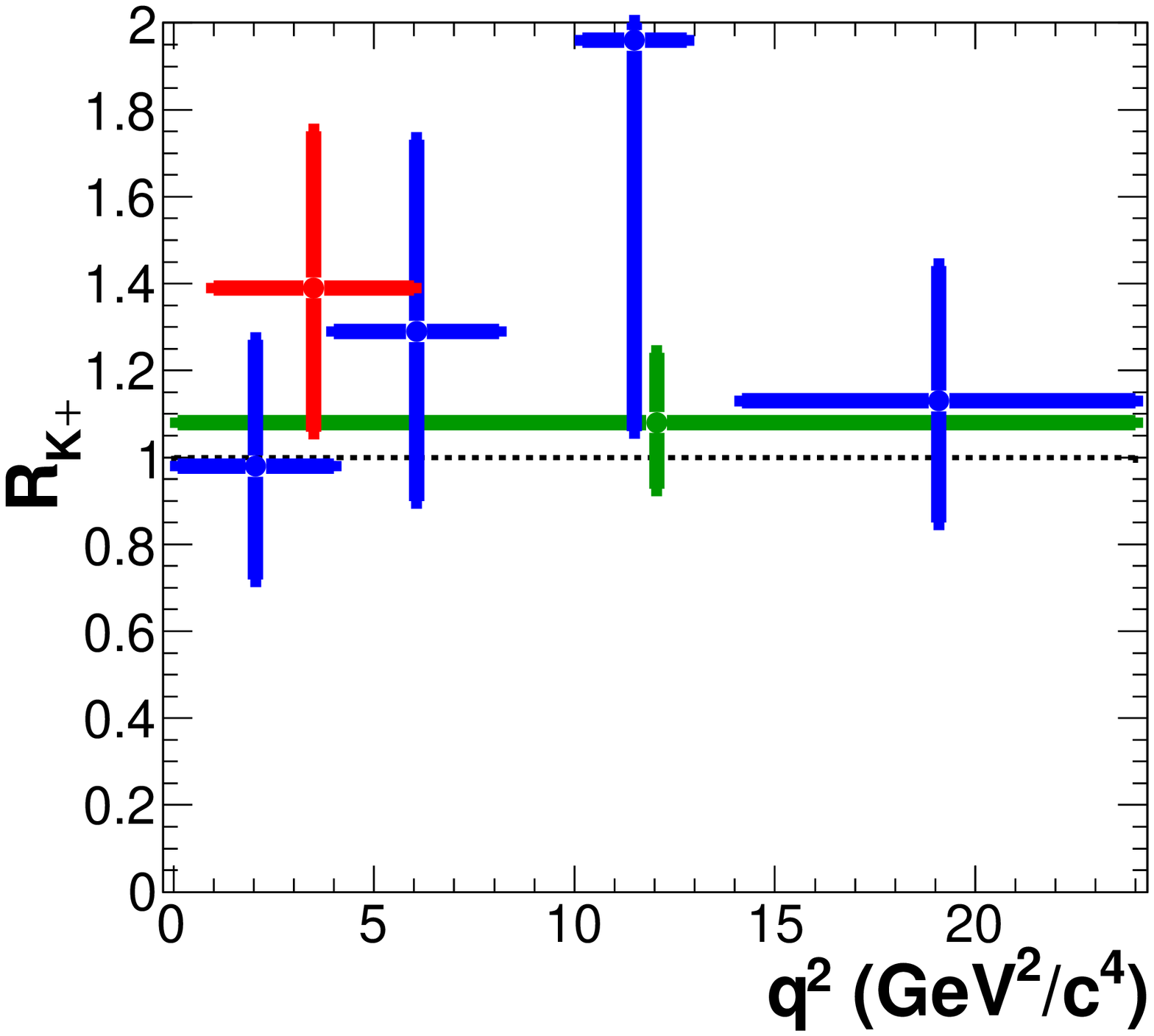}
    \includegraphics[width=0.5\columnwidth]{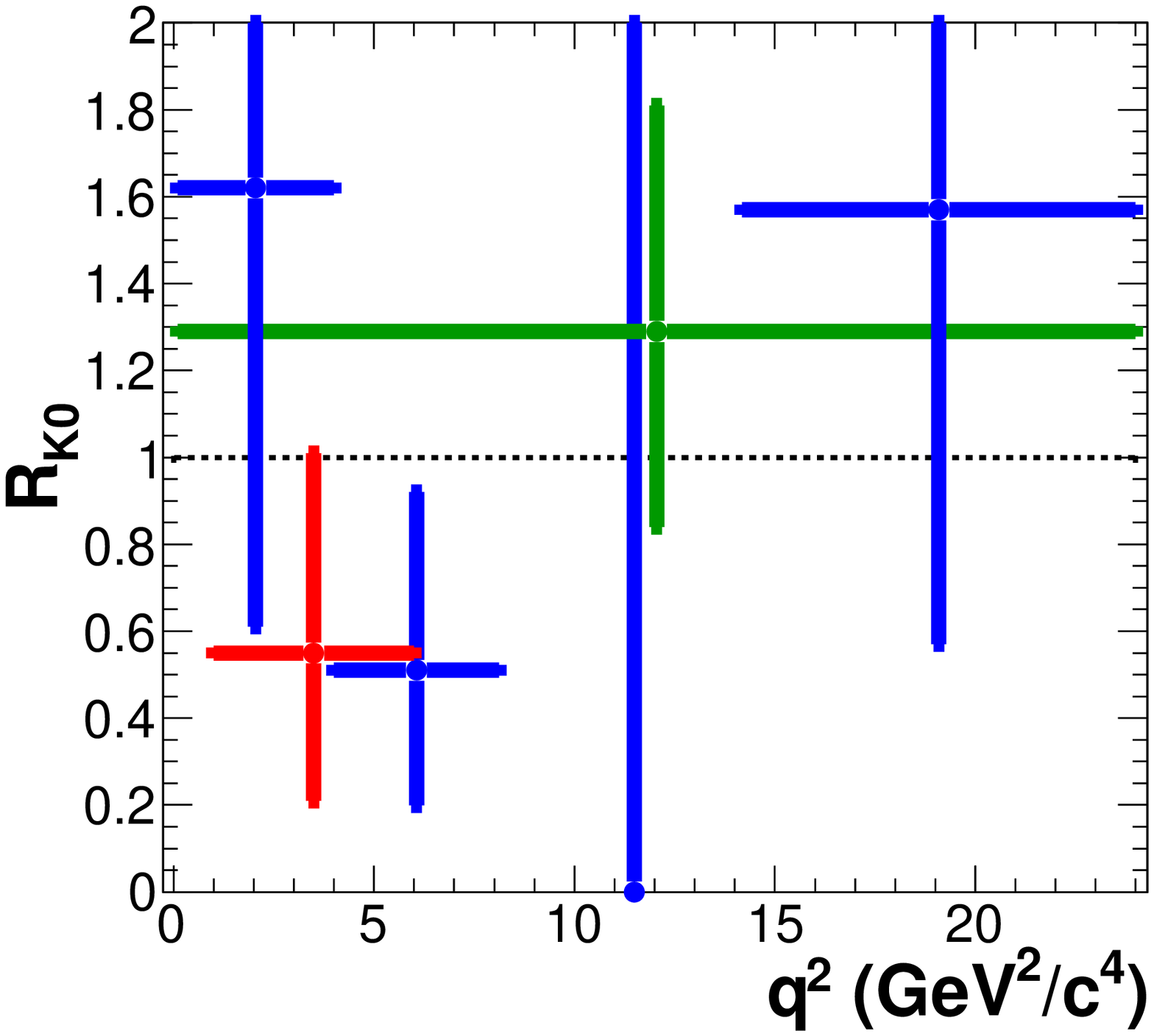}
    }
\centering
  \mbox{
    \includegraphics[width=0.75\columnwidth]{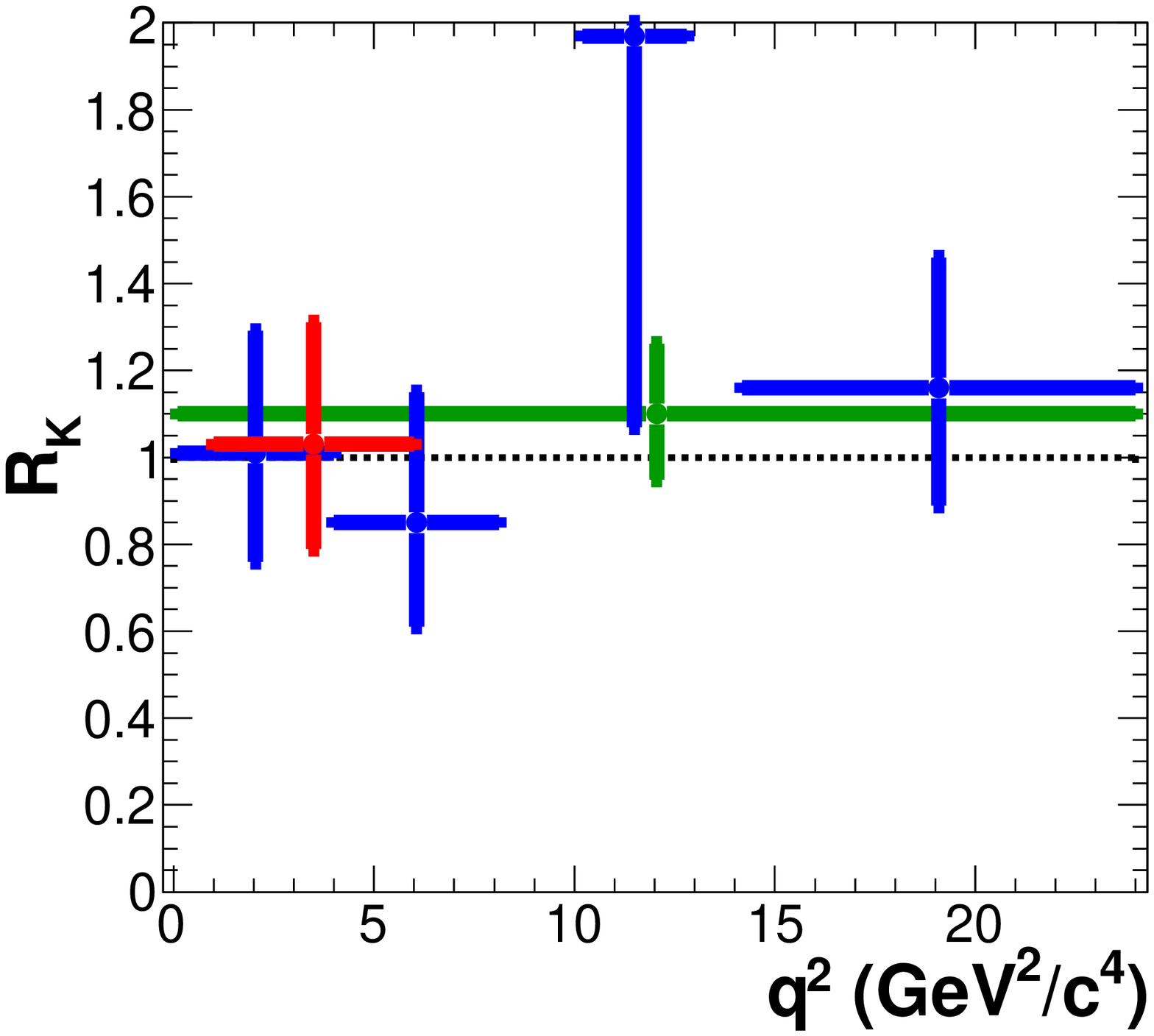}
  }
  \caption{$R_{K}$ in bins of $q^{2}$, for $B^{+}\to K^{+} \ell^{+} \ell^{-}$ (top-left), $B^{0}\to \KS \ell^{+} \ell^{-}$ (top-right), and both modes combined (bottom). The red marker represents the bin of $1.0<q^{2}<6.0$ \qq, and the blue markers are for $0.1<q^{2}<4.0$, $4.00<q^{2}<8.12$, $10.2<q^{2}<12.8$ and $q^{2}>14.18$ \qq~bins. The green marker denotes the whole $q^{2}$ region excluding the charmonium resonances.}
  \label{fig:RK}
\end{figure}

\begin{figure}[htbp]
  \mbox{
    \includegraphics[width=0.5\columnwidth]{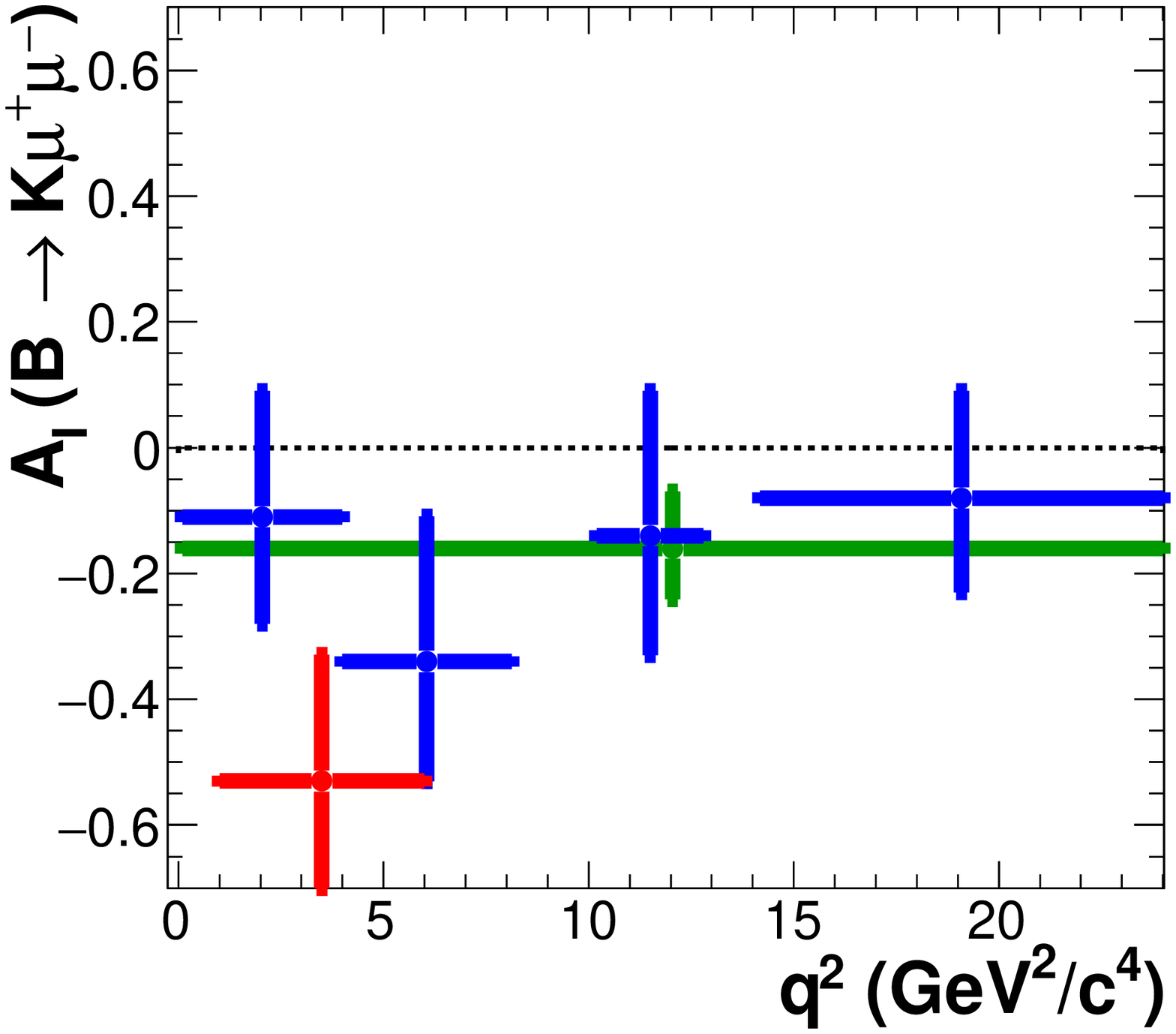}
    \includegraphics[width=0.5\columnwidth]{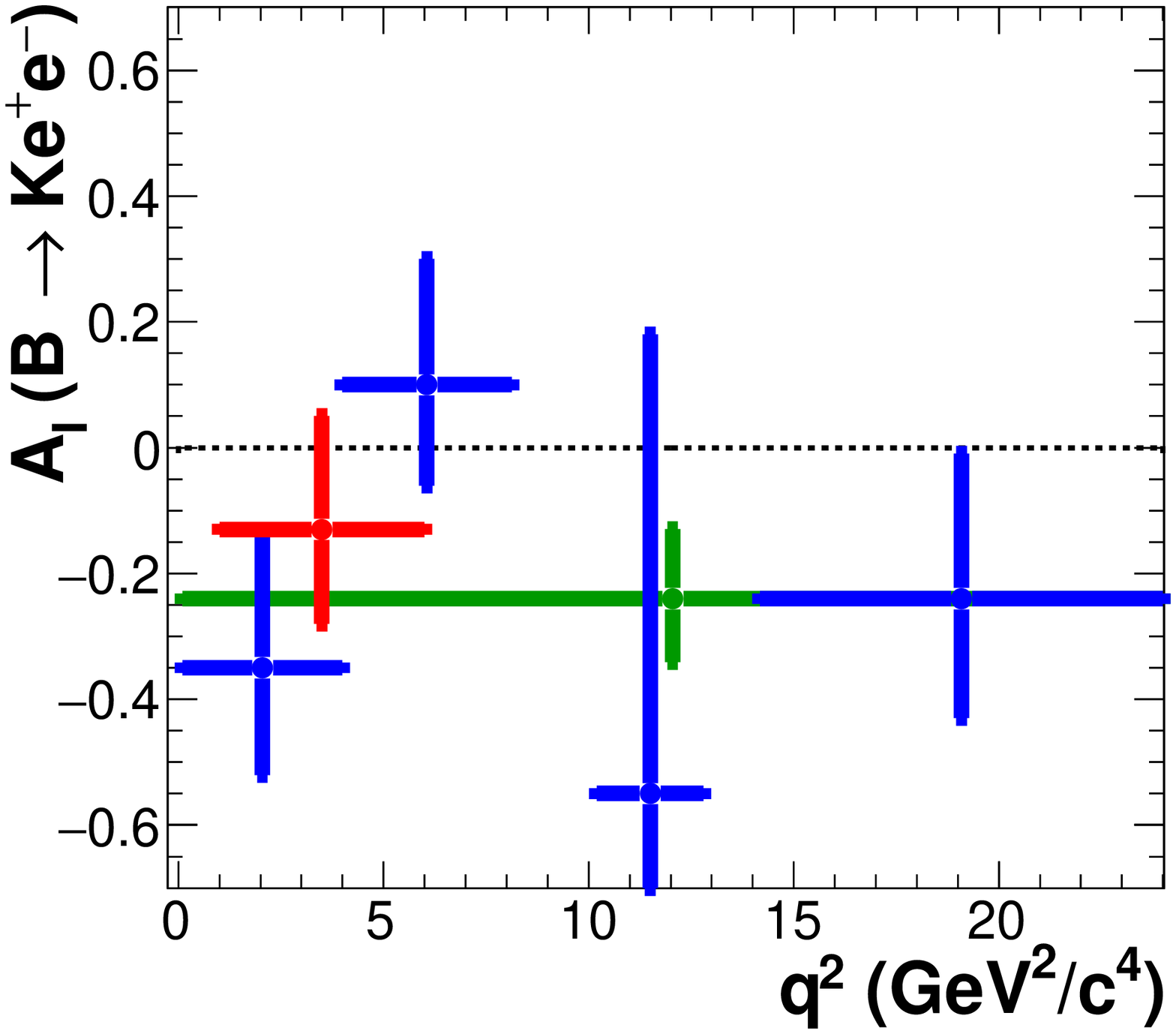}
    }

 \centering
 \mbox{
    \includegraphics[width=0.75\columnwidth]{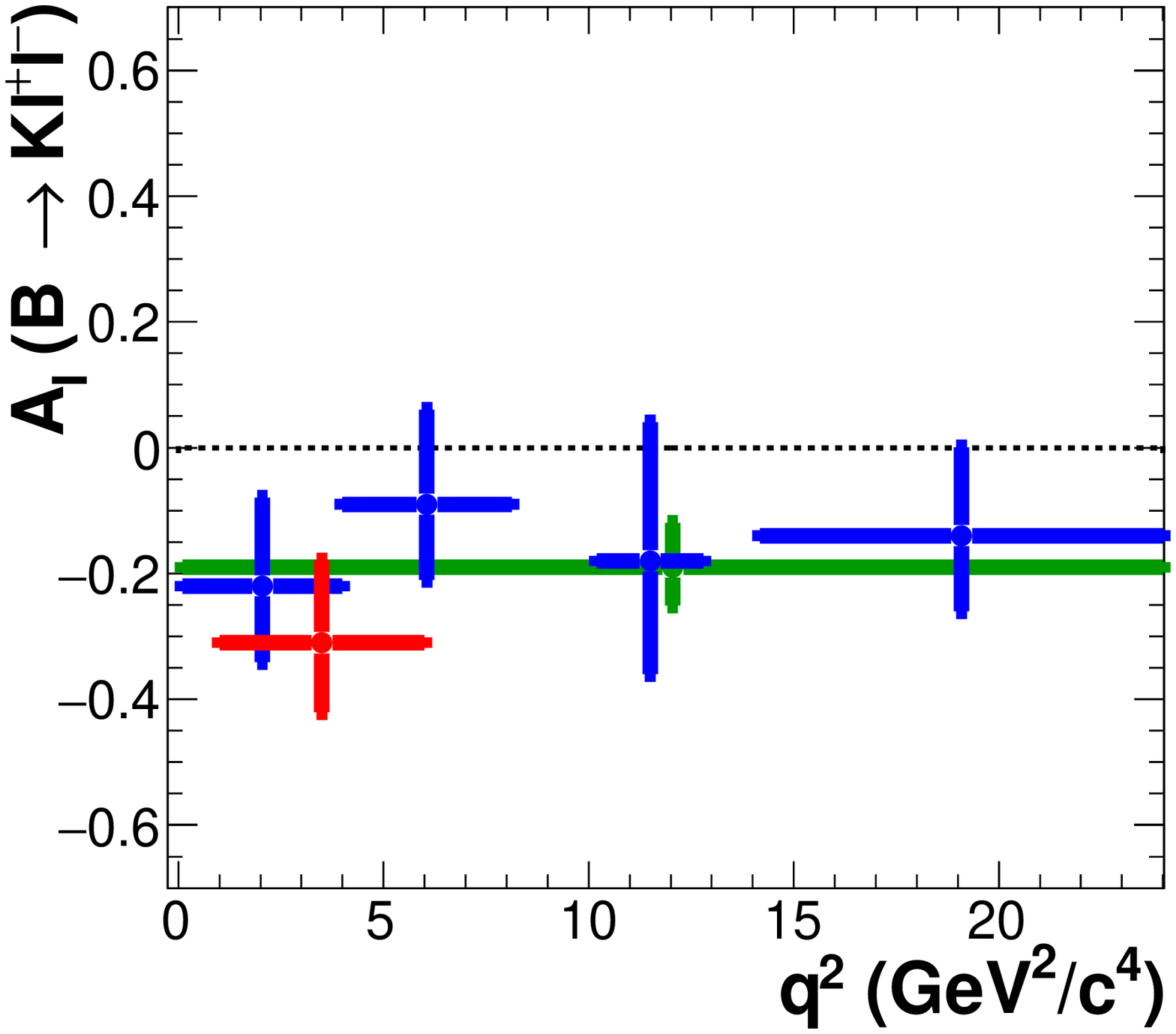}
  }
  \caption{$A_{I}$ measurements in bins of $q^{2}$, for decays $B \to K \mu^{+} \mu^{-}$ (top-left), $B \to K e^{+} e^{-}$ (top-right), and both modes combined (bottom). The legends are the same as in Fig. \ref{fig:RK}.}
  \label{fig:AI}
\end{figure}

\begin{figure}[htbp]
  \mbox{
    \includegraphics[width=0.5\columnwidth]{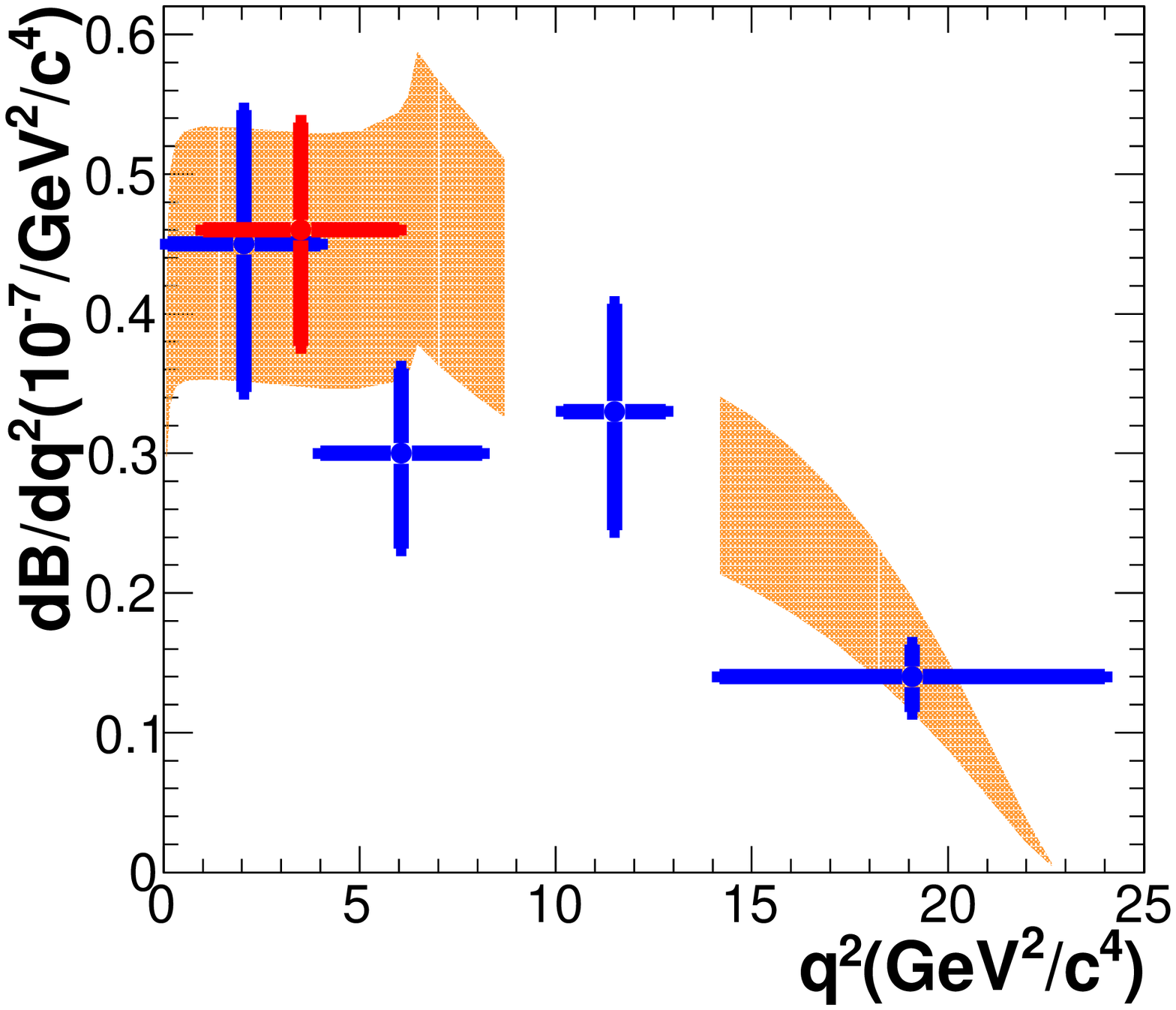}
    \includegraphics[width=0.5\columnwidth]{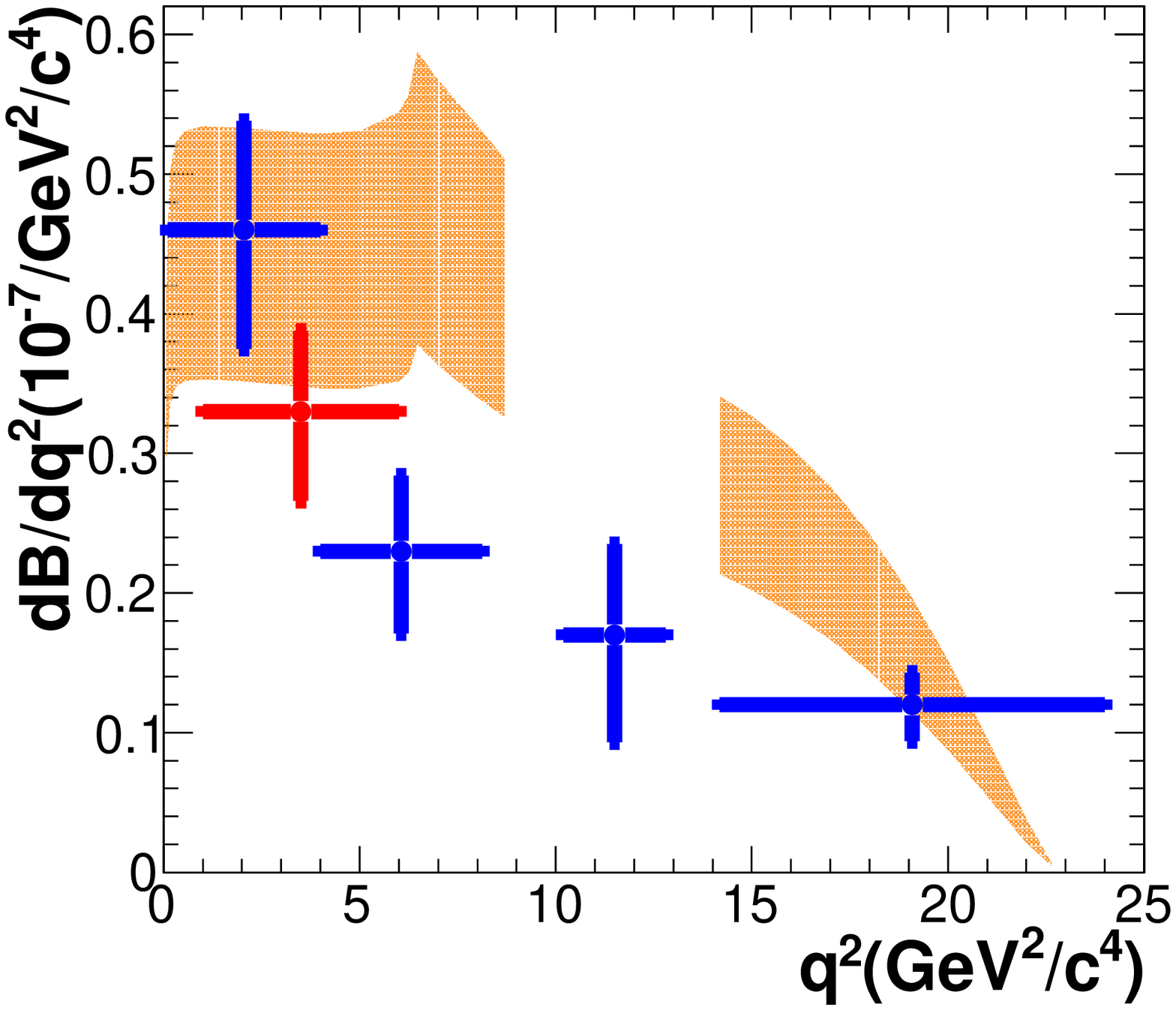}
    }
      \mbox{
    \includegraphics[width=0.5\columnwidth]{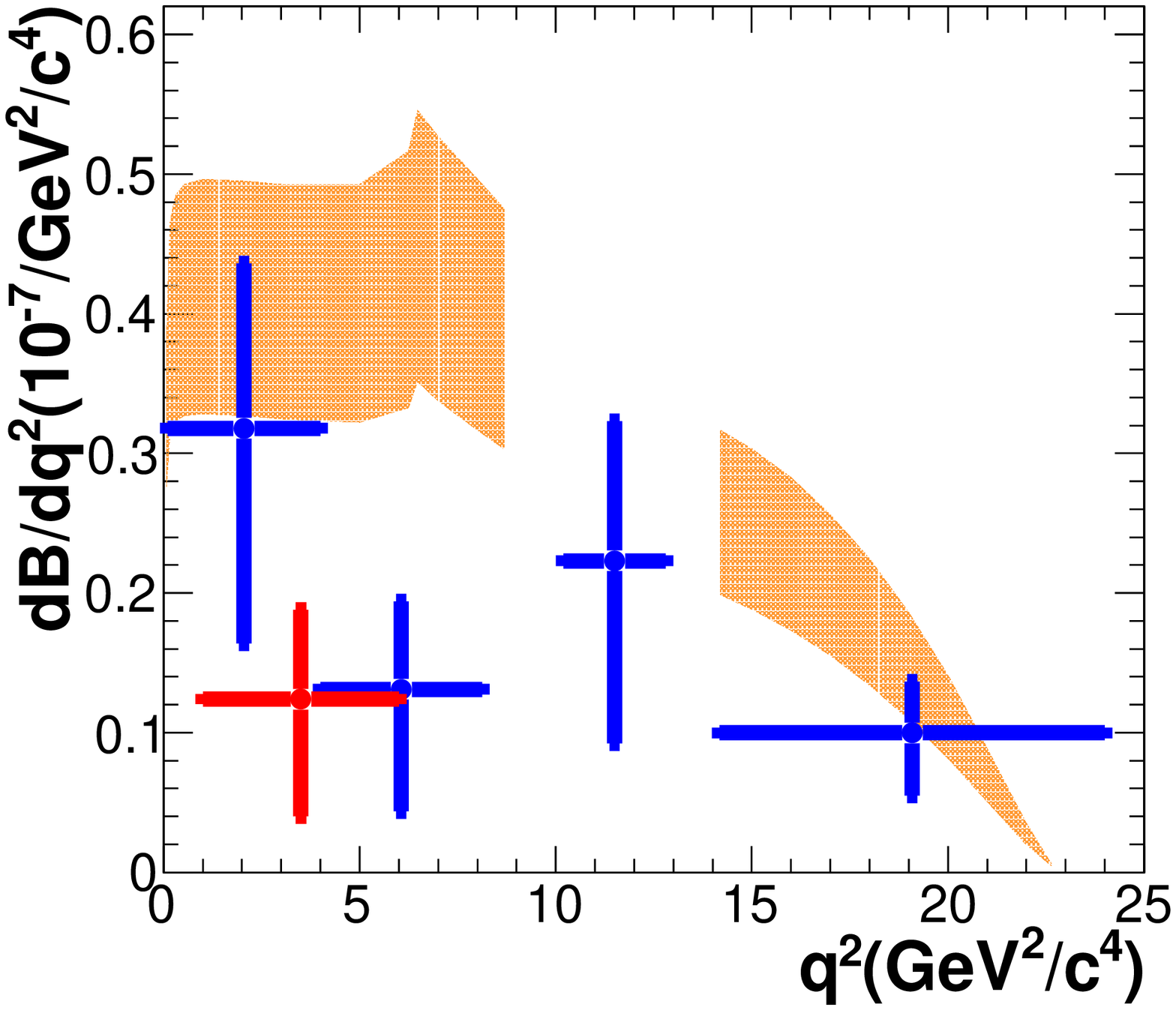}
    \includegraphics[width=0.5\columnwidth]{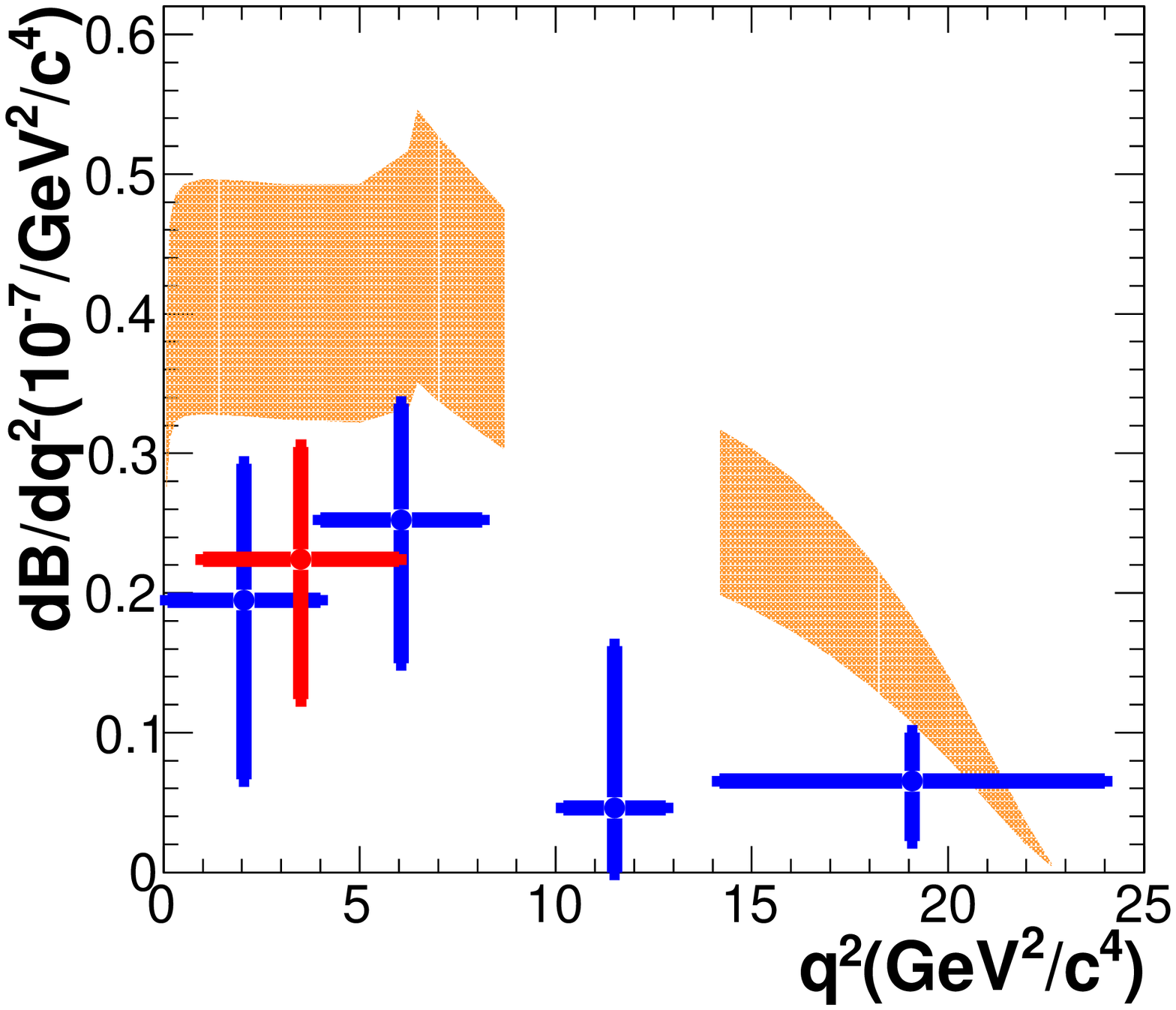}
  }
  \caption{${d\cal B}/{dq^{2}}$ measurements in bins of $q^{2}$, for decays $B^{+} \to K^{+} \mu^{+} \mu^{-}$ (top-left), $B^{+} \to K^{+} e^{+} e^{-}$ (top-right), $B^{0} \to K^{0} \mu^{+} \mu^{-}$ (bottom-left), and $B^{0} \to K^{0} e^{+} e^{-}$ (bottom-right). The legends are the same as in Fig. \ref{fig:RK}. The yellow shaded regions show the theoretical predictions from the light-cone sum rule and lattice QCD calculations \cite{brtheory1,brtheory2}.}
  \label{fig:BR}
\end{figure}

\begin{figure}[htbp]
  \mbox{
    \includegraphics[width=0.33\columnwidth]{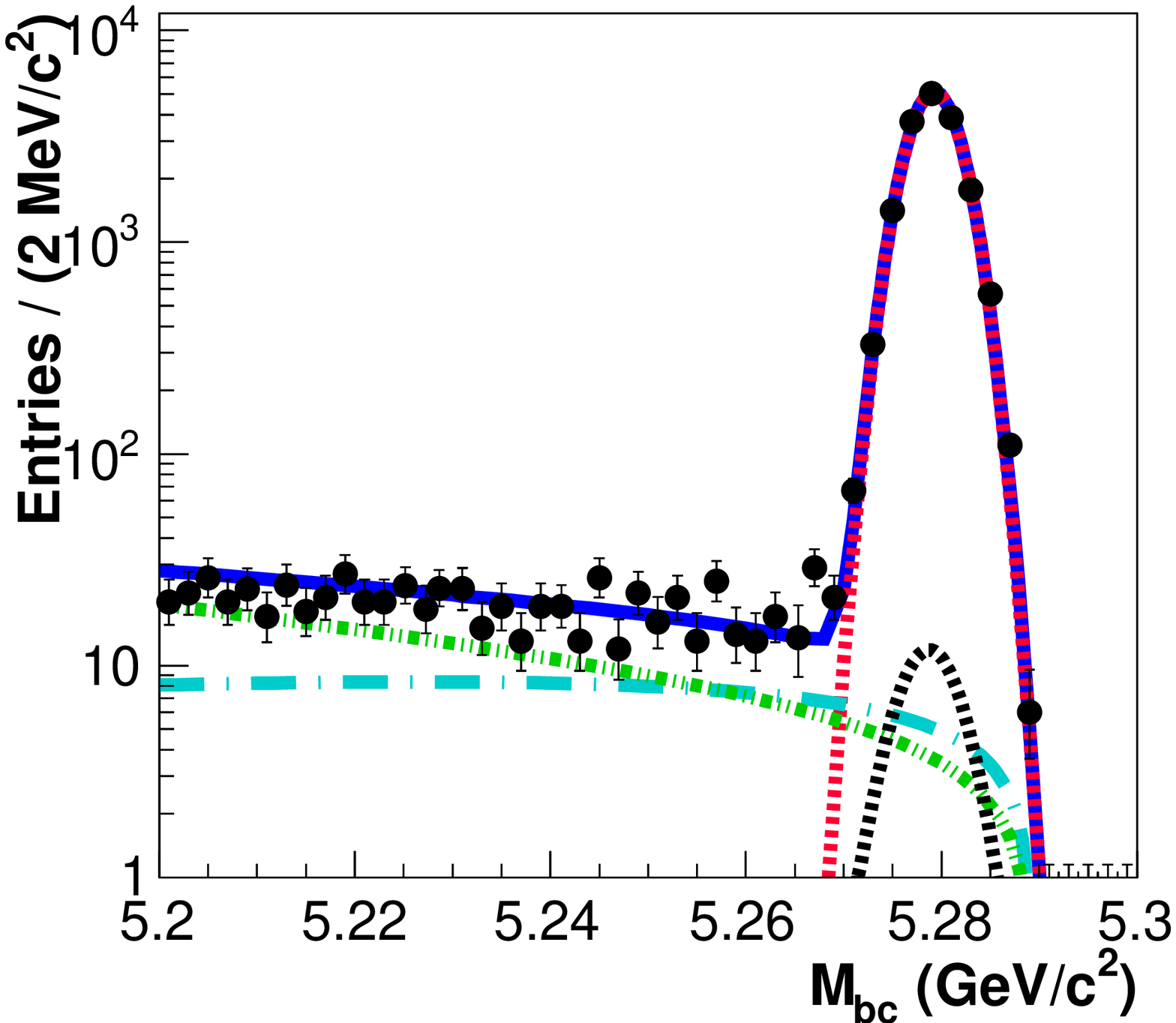}
    \includegraphics[width=0.33\columnwidth]{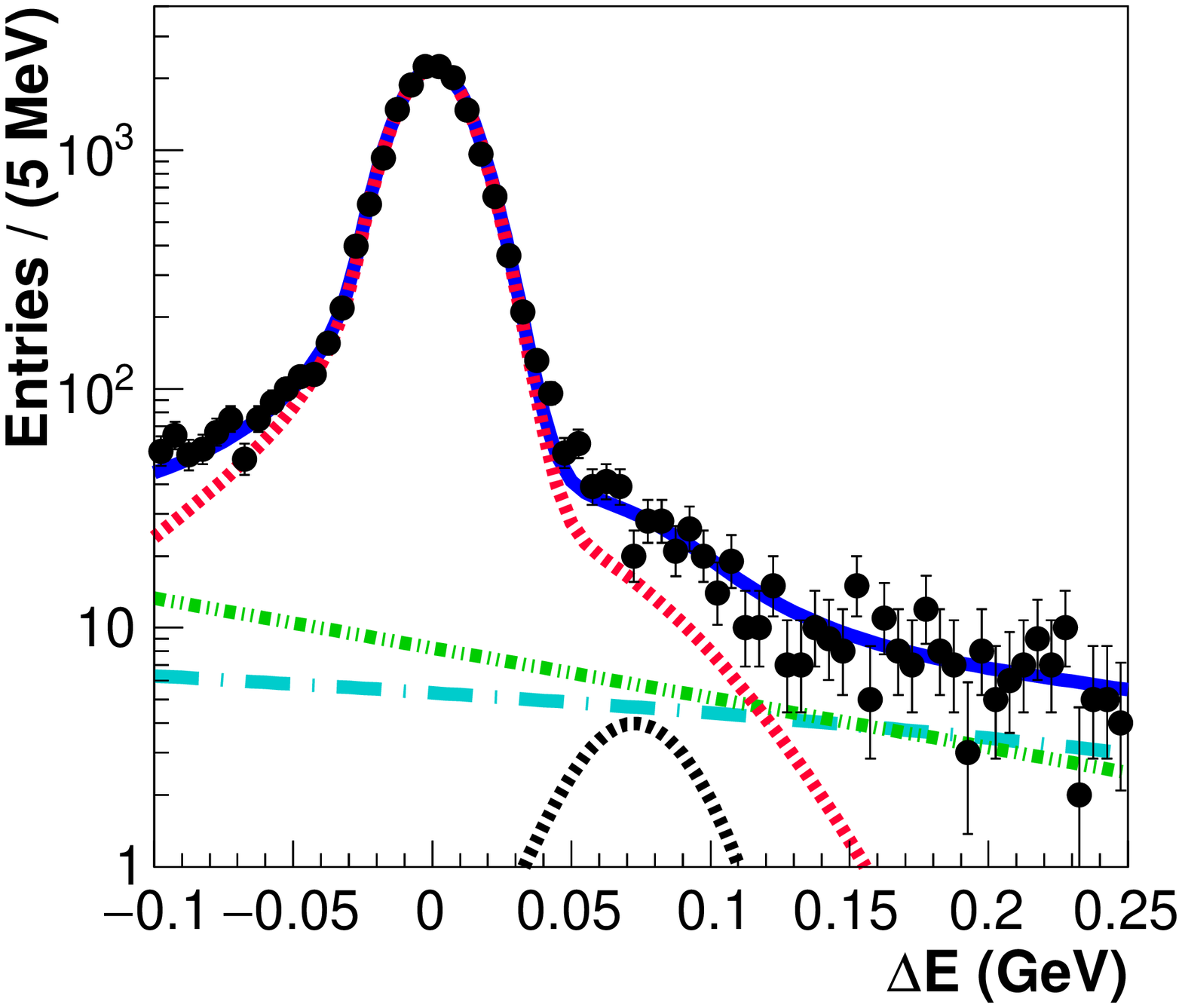}
    \includegraphics[width=0.33\columnwidth]{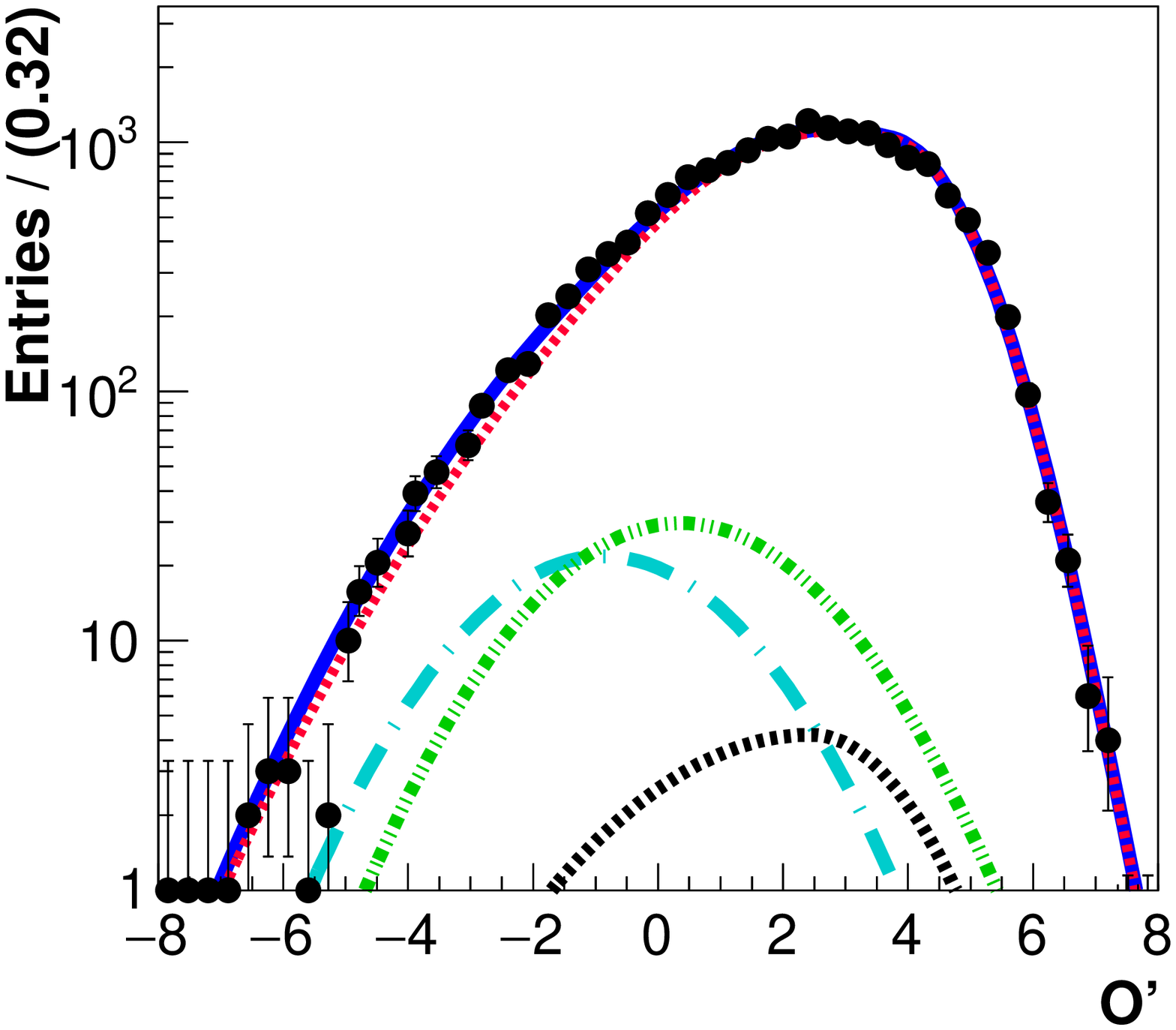}
    }
  \mbox{
    \includegraphics[width=0.33\columnwidth]{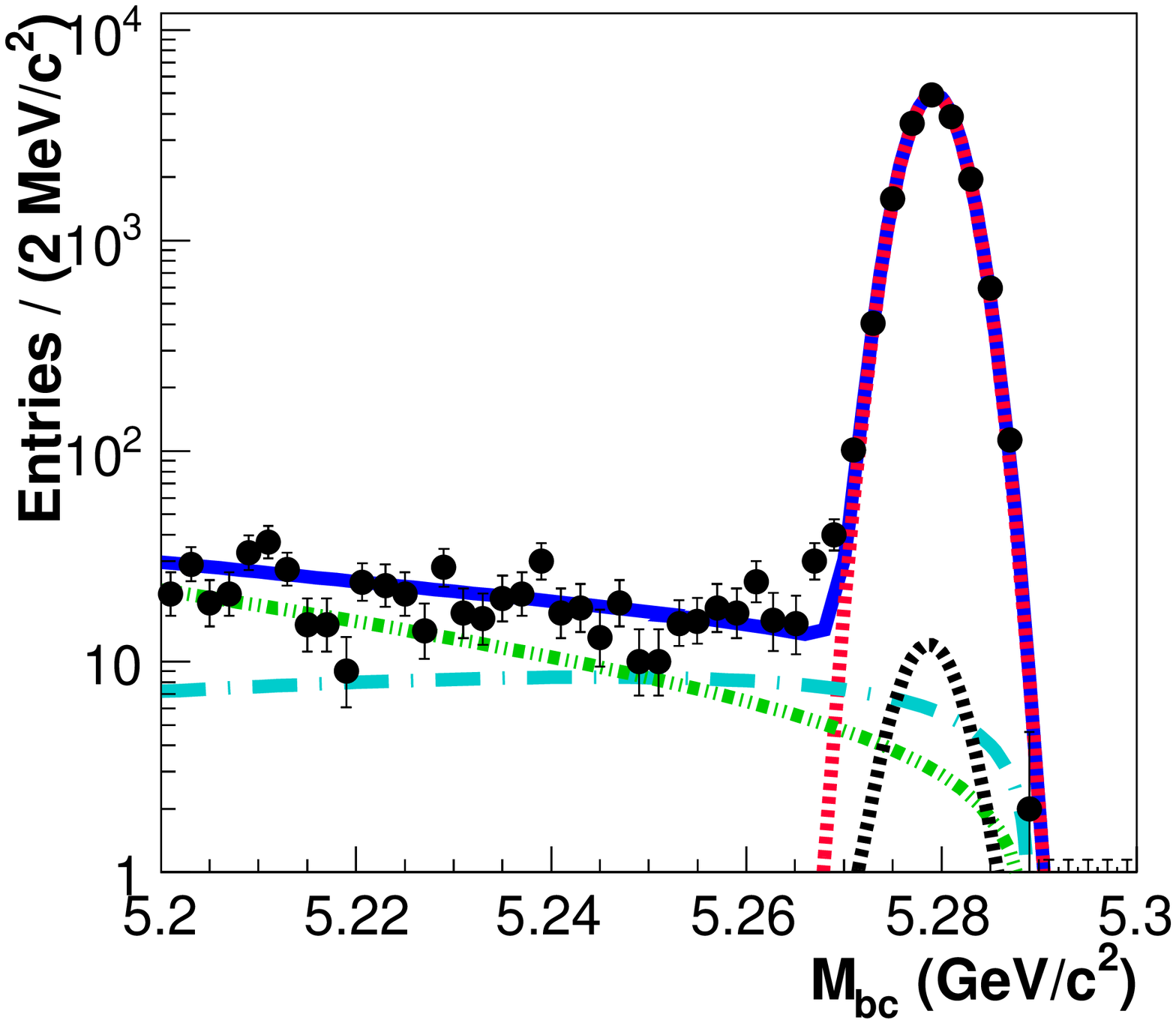}
    \includegraphics[width=0.33\columnwidth]{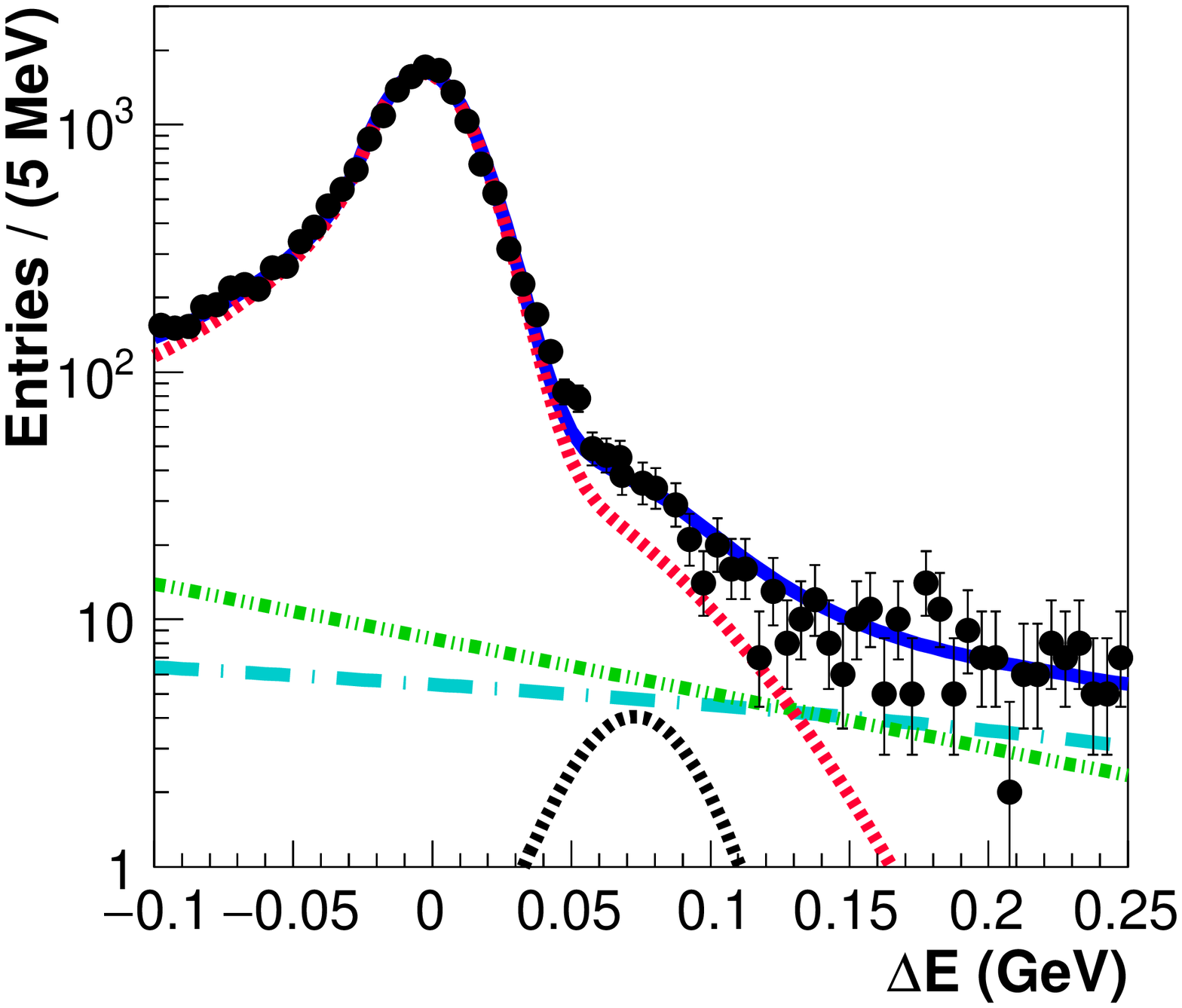}
    \includegraphics[width=0.33\columnwidth]{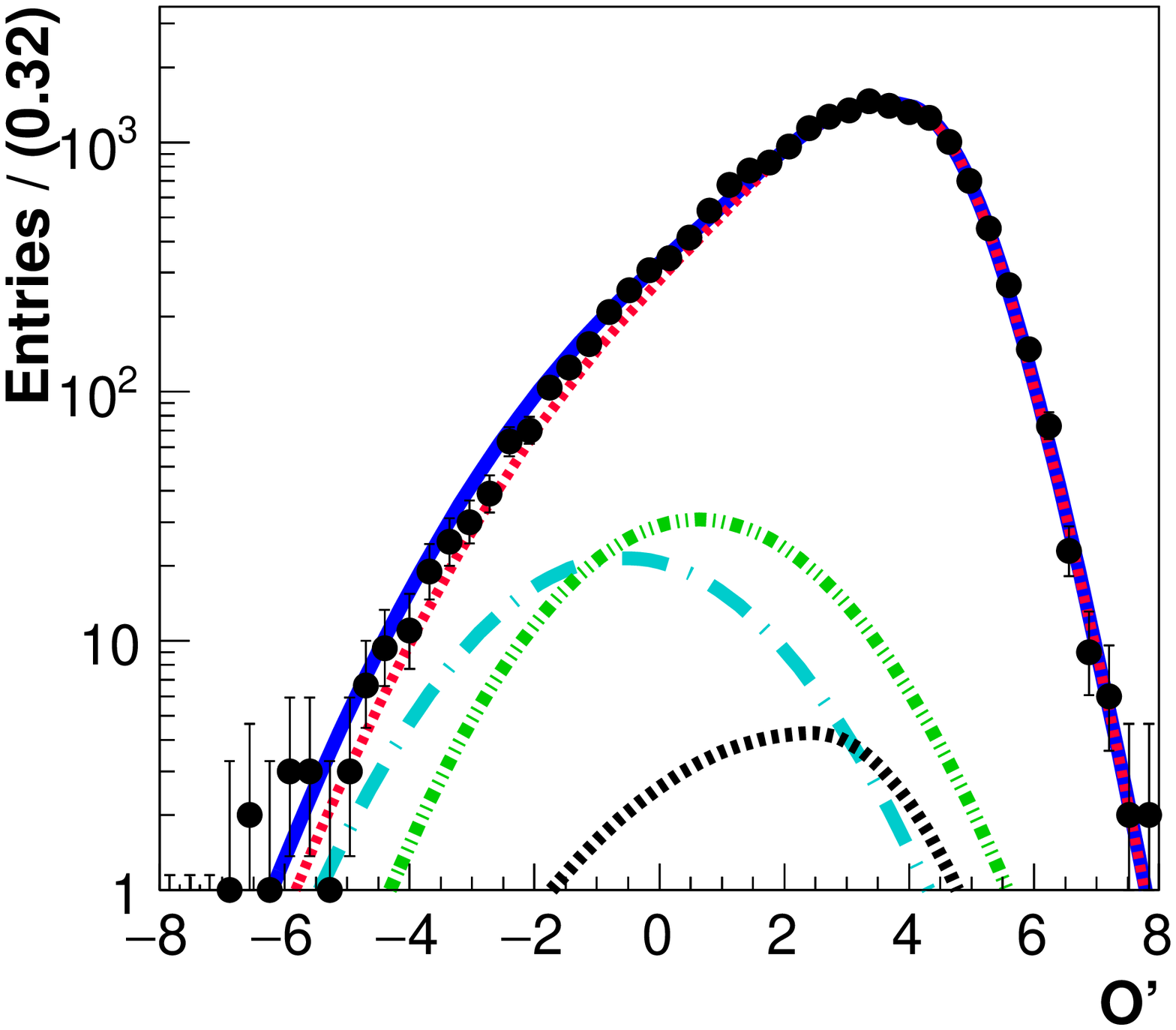}
    }
  \caption{\Mbc~(left), $\Delta E$ (middle), and ${\cal O'}$ (right) projections of three-dimensional unbinned extended maximum-likelihood fits to the data events that pass the selection criteria for $B^{+} \to J/\psi (\rightarrow \mu^{+}\mu^{-}) K^{+}$ (top), and $B^{+} \to J/\psi (\rightarrow e^{+}e^{-}) K^{+}$ (bottom). The legends are the same as in Fig. \ref{fig:btokpll} and black dashed curve is $[\pi^{+} J/\psi]$ background.}
  \label{fig:btojpsik}
\end{figure}

There are $137 \pm 14$ and $138 \pm 15$ signal events for the decays $B^{+}\rightarrow K^{+}\mu^{+}\mu^{-}$ and $B^{+}\rightarrow K^{+}e^{+}e^{-}$, respectively, whereas the yields for the decays $B^{0}\rightarrow \KS\mu^{+}\mu^{-}$ and $B^{0}\rightarrow \KS e^{+}e^{-}$ are $27.3^{+6.6}_{-5.8}$ and $21.8^{+7.0}_{-6.1}$ events, respectively. The fit is also performed in the aforementioned five $q^{2}$ bins [(0.1, 4.0), (4.00, 8.12), (1.0, 6.0), (10.2, 12.8), and ($>14.18$)]\qq~including the $(1.0,6.0)$\qq~bin, where LHCb reports a possible deviation in $R_{K^{+}}$, and $R_{K}$ and $A_{I}$ values are calculated from Eqs.~(\ref{RK}) and~(\ref{AI}), respectively. 
The results are listed in Table~\ref{tab:lfu} and $R_{K}$ and $A_{I}$ are also shown in Figs.~\ref{fig:RK} and~\ref{fig:AI}, respectively. 
The differential branching fraction (${d\cal B}/{dq^{2}}$) results are shown in Fig.~\ref{fig:BR}. 
The branching fractions for the $B^{+} \rightarrow K^{+}\ell^{+}\ell^{-}$, and $B^{0} \rightarrow K^{0}\ell^{+}\ell^{-}$ modes are $(5.99^{+0.45}_{-0.43} \pm 0.14)\times 10^{-7}$, and $(3.51^{+0.69}_{-0.60}\pm 0.10) \times 10^{-7}$, respectively for the whole $q^{2}$ range. The measurement is done for $B^{0} \rightarrow \KS\ell^{+}\ell^{-}$, but the branching fraction is quoted for $B^{0} \rightarrow K^{0}\ell^{+}\ell^{-}$, considering a factor of 2. 
Figure~\ref{fig:btojpsik} illustrates the fit for $B \rightarrow  J/\psi(\rightarrow \ell^+\ell^-) K$ modes and the corresponding branching fractions obtained are listed in Table~\ref{tab:btokjpsi}. These samples serve as calibration modes for the PDF shapes used as well as to calibrate the efficiency of $\cal O$~$> \cal O_{\rm{min}}$ requirement for possible difference between data and simulation. These are also used to verify that there is no bias for some of the key observables. For example, we obtain $R_K(J/\psi) = 0.994 \pm 0.011 \pm 0.010$ and $0.993 \pm 0.015 \pm 0.010$ for $B^+ \rightarrow J/\psi K^+$ and $B \rightarrow J/\psi \KS$, respectively. Similarly, $A_{I} (B \rightarrow J/\psi K)$ is $-0.002 \pm 0.006 \pm 0.014$.
\begin{landscape}
%\begingroup
%\squeezetable
\begin{table*}[p]
%  \renewcommand{\arraystretch}{1.5}
 % \begin{ruledtabular}
    \caption{Results from the fits. The columns correspond to the $q^{2}$ bin size, decay mode, reconstruction efficiency, signal yield, branching fraction, lepton-flavor-separated and combined $A_{I}$ and $R_{K}$.}
    \label{tab:lfu}
    %\begin{adjustbox}{angle=90}
    
    \begin{tabular}{clccccccc}
      $q^{2}$ & $B \to $ mode & $\varepsilon$ & $N^{}_{\rm sig}$ & ${\cal B}$ & $A_{I}$ & $A_{I}$ & $R_{K}$ & $R_{K}$  \\
      (\qq) &  & (\%) &  &  $(10^{-7})$ &(individual)  & (combined) &(individual) & (combined)\\
      \hline
      \multirow{4}{*}{(0.1,4.0)} & $K^{+} \mu^{+} \mu^{-}$ & $20.4$ & $28.4^{+6.6}_{-5.9}$ & $1.76^{+0.41}_{-0.37} \pm 0.04$ & $A_{I}(\mu\mu)=$ & \multirow{4}{*}{$-0.22^{+0.14}_{-0.12} \pm 0.01$} & $R_{K^+} =$ & \multirow{4}{*}{$1.01^{+0.28}_{-0.25} \pm 0.02$} \\
      & $\KS \mu^{+} \mu^{-}$ & $14.7$ & $6.8^{+3.3}_{-2.6}$ & $0.62^{+0.30}_{-0.23} \pm 0.02$ & $-0.11^{+0.20}_{-0.17} \pm 0.01$ & & $ 0.98^{+0.29}_{-0.26}\pm0.02$ & \\
      & $K^{+} e^{+} e^{-}$ & $29.1$ & $41.5^{+7.7}_{-7.0}$ & $1.80^{+0.33}_{-0.30} \pm 0.05$ & $A_{I}(ee)=$ &  & $R_{\KS} =$ & \\
      & $\KS e^{+} e^{-}$ & $19.3$ & $5.5^{+3.6}_{-2.7}$ & $0.38^{+0.25}_{-0.19} \pm 0.01$ & $-0.35^{+0.21}_{-0.17} \pm 0.01$ & & $1.62^{+1.31}_{-1.01}\pm0.02$ & \\
      \hline
      \multirow{4}{*}{(4.00,8.12)} & $K^{+} \mu^{+} \mu^{-}$ & $29.0$ & $28.4^{+6.4}_{-5.7}$ & $1.24^{+0.28}_{-0.25} \pm 0.03$ & $A_{I}(\mu\mu)=$ & \multirow{4}{*}{$-0.09^{+0.15}_{-0.12} \pm 0.01$} & $R_{K^+} =$ & \multirow{4}{*}{$0.85^{+0.30}_{-0.24} \pm 0.01$} \\
      & $\KS \mu^{+} \mu^{-}$ & $21.0$ & $4.2^{+4.2}_{-3.5}$ & $0.27^{+0.18}_{-0.13} \pm 0.01$ & $-0.34^{+0.23}_{-0.19} \pm 0.01$ & & $1.29^{+0.44}_{-0.39}\pm0.02$ & \\
      & $K^{+} e^{+} e^{-}$ & $35.4$ & $26.9^{+6.9}_{-6.1}$ & $0.96^{+0.24}_{-0.22} \pm 0.03$ & $A_{I}(ee)=$ &  & $R_{\KS} =$ & \\
      & $\KS e^{+} e^{-}$ & $23.9$ & $9.3^{+3.7}_{-3.0}$ & $0.52^{+0.21}_{-0.17} \pm 0.02$ & $0.10^{+0.20}_{-0.16}\pm 0.01$ & & $0.51^{+0.41}_{-0.31}\pm0.01$ & \\
      \hline
      \multirow{4}{*}{(1.0,6.0)} & $K^{+} \mu^{+} \mu^{-}$ & $23.2$ & $42.3^{+7.6}_{-6.9}$ & $2.30^{+0.41}_{-0.38} \pm 0.05$ & $A_{I}(\mu\mu)=$ & \multirow{4}{*}{$-0.31^{+0.13}_{-0.11} \pm 0.01$} & $R_{K^+} =$ & \multirow{4}{*}{$1.03^{+0.28}_{-0.24}\pm 0.01$} \\
      & $\KS \mu^{+} \mu^{-}$ & $16.8$ & $3.9^{+2.7}_{-2.0}$ & $0.31^{+0.22}_{-0.16} \pm 0.01$ & $-0.53^{+0.20}_{-0.17} \pm 0.02$ & & $1.39^{+0.36}_{-0.33}\pm 0.02$ & \\
      & $K^{+} e^{+} e^{-}$ & $31.7$ & $41.7^{+8.0}_{-7.2}$ & $1.66^{+0.32}_{-0.29} \pm 0.04$ & $A_{I}(ee)=$ &  & $R_{\KS} =$ & \\
      & $\KS e^{+} e^{-}$ & $21.1$ & $8.9^{+4.0}_{-3.2}$ & $0.56^{+0.25}_{-0.20} \pm 0.02$ & $-0.13^{+0.18}_{-0.15} \pm 0.01$ & & $0.55^{+0.46}_{-0.34}\pm 0.01$ & \\
      \hline
      \multirow{4}{*}{(10.2,12.8)} & $K^{+} \mu^{+} \mu^{-}$ & $35.6$ & $24.3^{+6.3}_{-5.5}$ & $0.86^{+0.22}_{-0.20} \pm 0.02$ & $A_{I}(\mu\mu)=$ & \multirow{4}{*}{$-0.18^{+0.22}_{-0.18} \pm 0.01$} & $R_{K^+} =$ & \multirow{4}{*}{$1.97^{+1.03}_{-0.89}\pm 0.02$} \\
      & $\KS \mu^{+} \mu^{-}$ & $26.5$ & $5.7^{+3.4}_{-2.6}$ & $0.29^{+0.17}_{-0.13} \pm 0.01$ & $-0.14^{+0.24}_{-0.19} \pm 0.01$ & & $1.96^{+1.03}_{-0.89}\pm 0.02$ & \\
      & $K^{+} e^{+} e^{-}$ & $40.3$ & $14.0^{+6.4}_{-5.5}$ & $0.44^{+0.20}_{-0.17} \pm 0.01$ & $A_{I}(ee)=$ &  & $R_{\KS} =$ & \\
      & $\KS e^{+} e^{-}$ & $26.5$ & $1.1^{+3.7}_{-3.0}$ & $0.06^{+0.19}_{-0.15} \pm 0.01$ & $-0.55^{+0.73}_{-0.60} \pm 0.01$ & & $5.18^{+17.69}_{-14.32}\pm 0.06$ & \\
      \hline
      \multirow{4}{*}{$>14.18$} & $K^{+} \mu^{+} \mu^{-}$ &$45.2$ & $47.9^{+8.6}_{-7.8}$ & $1.34^{+0.24}_{-0.22} \pm 0.03$ & $A_{I}(\mu\mu)=$ & \multirow{4}{*}{$-0.14^{+0.14}_{-0.12} \pm 0.01$} & $R_{K^+} =$ & \multirow{4}{*}{$1.16^{+0.30}_{-0.27} \pm 0.01$} \\
      & $\KS \mu^{+} \mu^{-}$ & $25.7$ & $9.6^{+4.2}_{-3.5}$ & $0.49^{+0.22}_{-0.18} \pm 0.01$ & $-0.08^{+0.17}_{-0.15} \pm 0.01$ & & $1.13^{+0.31}_{-0.28}\pm 0.01$ & \\
      & $K^{+} e^{+} e^{-}$ & $46.2$ & $43.2^{+9.1}_{-8.3}$ & $1.18^{+0.25}_{-0.22} \pm 0.03$ & $A_{I}(ee)=$ &  & $R_{\KS} =$ & \\
      & $\KS e^{+} e^{-}$ & $24.9$ & $5.9^{+4.0}_{-3.1}$ & $0.32^{+0.21}_{-0.17} \pm 0.01$ & $-0.24^{+0.23}_{-0.19} \pm 0.01$ & & $1.57^{+1.28}_{-1.00}\pm 0.02$ & \\
      \hline
      \multirow{4}{*}{whole $q^{2}$} & $K^{+} \mu^{+} \mu^{-}$ & $27.8$ & $137.0^{+14.2}_{-13.5}$ & $6.24^{+0.65}_{-0.61} \pm 0.16$ & $A_{I}(\mu\mu)=$ & \multirow{4}{*}{$-0.19^{+0.07}_{-0.06} \pm 0.01$} & $R_{K^+} =$ & \multirow{4}{*}{$1.10^{+0.16}_{-0.15} \pm 0.02$} \\
      & $\KS \mu^{+} \mu^{-}$ & $18.5$ & $27.3^{+6.6}_{-5.9}$ & $1.97^{+0.48}_{-0.42} \pm 0.06$ & $-0.16^{+0.09}_{-0.08} \pm 0.01$ & & $1.08^{+0.16}_{-0.15}\pm0.02$ & \\
      & $K^{+} e^{+} e^{-}$ & $30.3$ & $138.0^{+15.5}_{-14.7}$ & $5.75^{+0.64}_{-0.61} \pm 0.15$ & $A_{I}(ee)=$ &  & $R_{\KS} =$ & \\
      & $\KS e^{+} e^{-}$ & $19.0$ & $21.8^{+7.0}_{-6.1}$ & $1.53^{+0.49}_{-0.43} \pm 0.04$ & $-0.24^{+0.11}_{-0.10} \pm 0.01$ & & $1.29^{+0.52}_{-0.45}\pm 0.01$ & \\
    \end{tabular}
%\end{adjustbox}
%  \end{ruledtabular}
\end{table*}
%\endgroup
\end{landscape}

\begin{table} 
\caption{Branching fraction for $B \rightarrow K\ell^{+}\ell^{-}$ and $B \rightarrow J/\psi K$ decays.}
\label{tab:btokjpsi}
\begin{center}
\begin{tabular}{lc}
  Mode &  $\cal B$  \\ \hline 
  $B^{+} \rightarrow K^{+}\ell^{+}\ell^{-}$ & $(5.99^{+0.45}_{-0.43}\pm 0.14) \times 10^{-7}$\\
$B^{0} \rightarrow K^{0}\ell^{+}\ell^{-}$ & $(3.51^{+0.69}_{-0.60}\pm 0.10) \times 10^{-7}$\\
\hline
$B^{+}\rightarrow J/\psi K^{+}$ & $(1.032 \pm 0.007 \pm 0.024) \times 10^{-3}$ \\
$B^{0}\rightarrow J/\psi K^{0}$ & $(0.902 \pm 0.010 \pm 0.026) \times 10^{-3}$ \\ 
\hline

\end{tabular}
\end{center}
\end{table}

\begin{figure}[htbp]
  \mbox{
    \includegraphics[width=0.33\columnwidth]{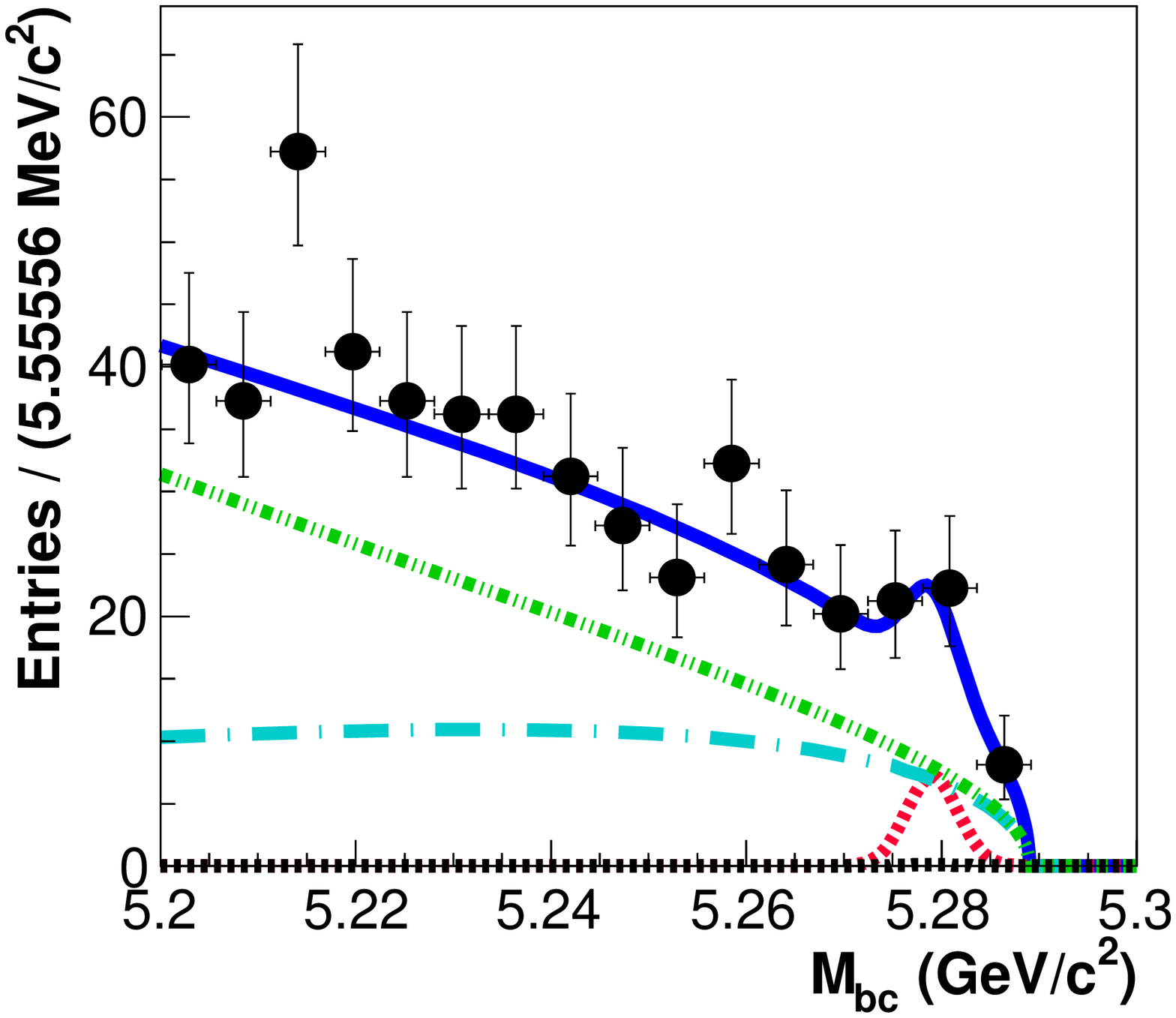}
    \includegraphics[width=0.33\columnwidth]{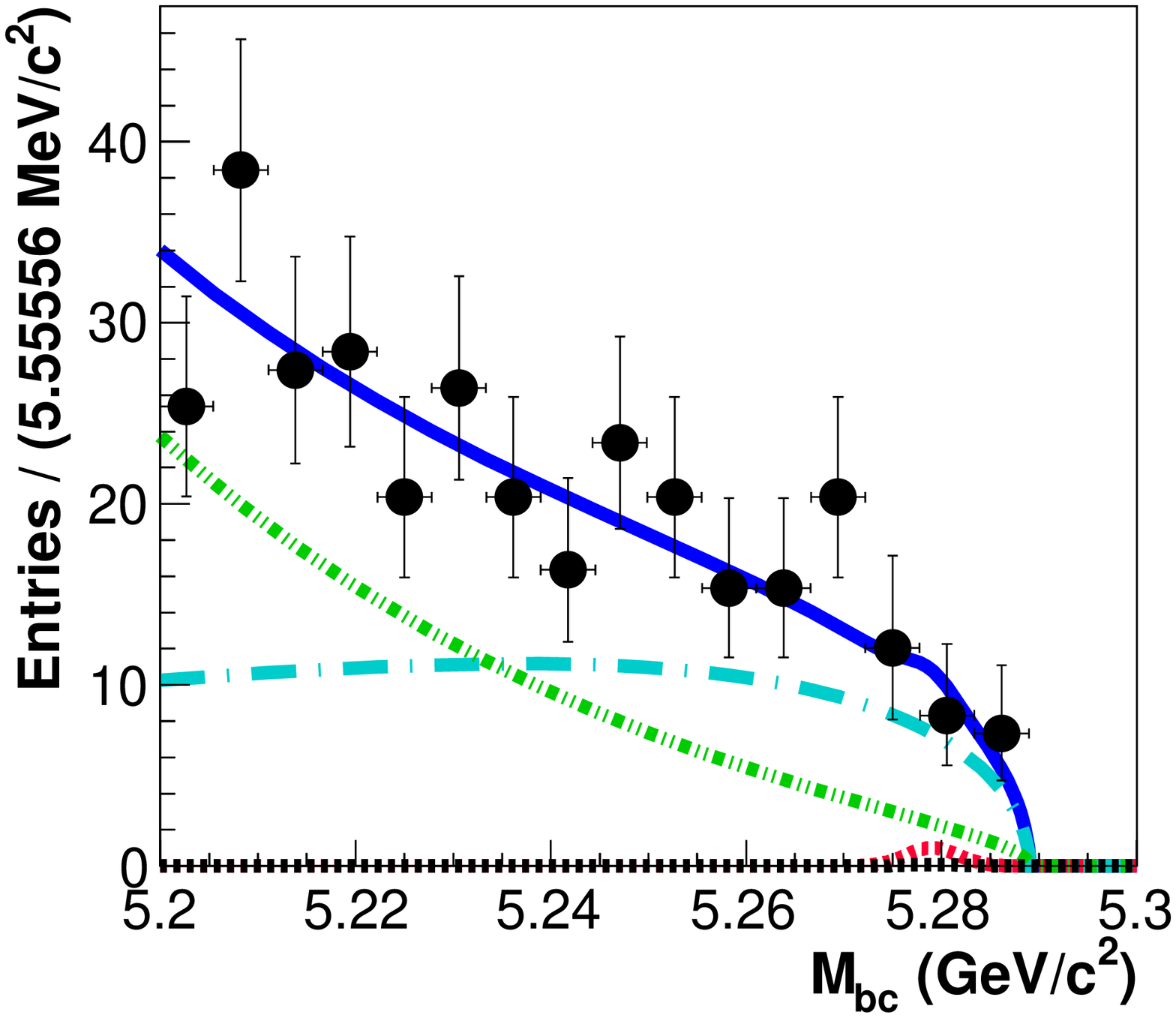}
    \includegraphics[width=0.33\columnwidth]{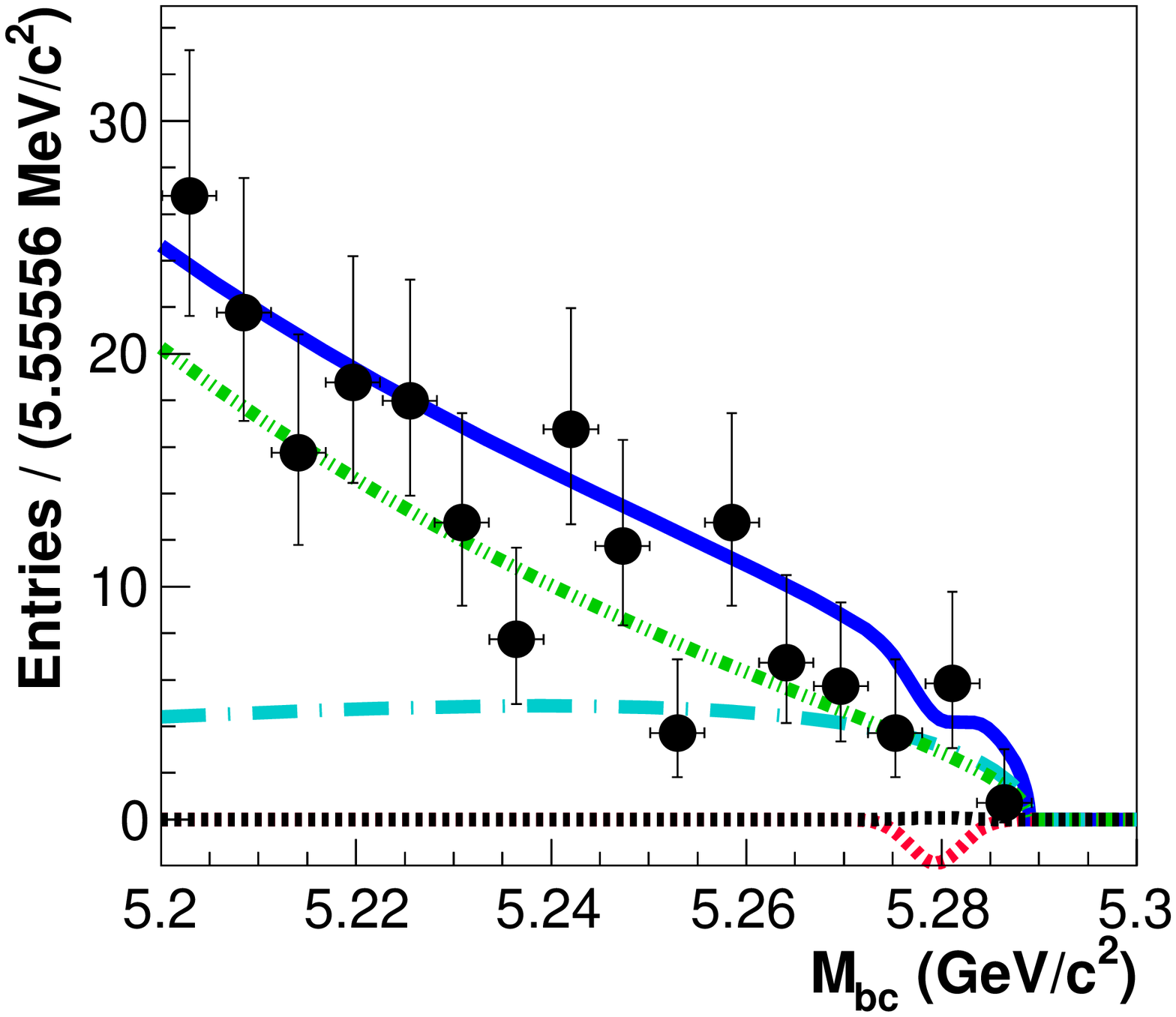}
    }
  \caption{Signal-enhanced \Mbc~projection of three-dimensional unbinned extended maximum-likelihood fits to the data events that pass the selection criteria for decays $B^{+} \rightarrow K^{+}\mu^{+}e^{-}$ (left), $B^{+} \rightarrow K^{+}\mu^{-}e^{+}$ (middle), and $B^{0} \rightarrow \KS\mu^{\pm}e^{\mp}$ (right). The legends are same as in Fig. \ref{fig:btokpll}.}
  \label{lfv:btokmue}
\end{figure}

The signal yields for LFV decays are obtained by performing unbinned extended maximum-likelihood fits, similar to those for the $B \rightarrow K\ell^{+}\ell^{-}$ modes. The signal-enhanced projection plots with fit results for LFV decays are shown in Fig.\ref{lfv:btokmue}.
The fitted yields are $11.6^{+6.1}_{-5.5}$, $1.7^{+3.6}_{-2.2}$, and $-3.3^{+4.0}_{-2.8}$ for $B^{+} \rightarrow K^{+}\mu^{+}e^{-}$, $B^{+} \rightarrow K^{+}\mu^{-}e^{+}$, and $B^{0} \rightarrow K_{S}^{0}\mu^{\pm}e^{\mp}$, respectively. 
For the $B^{0} \rightarrow K_{S}^{0}\mu^{\pm}e^{\mp}$ modes, we consider $\cal B(\rm\it{{{B^{{\rm{0}}} \rightarrow K_{S}^{{\rm{0}}} \mu^{+}e^{-}}}})$ and $\cal B(\rm\it{{{B^{{\rm{0}}} \rightarrow K_{S}^{{\rm{0}}} \mu^{-}e^{+}}}})$ together, as we do not distinguish between $B^{0}$ and ${\bar{B}^{0}}$. The total branching fraction $\cal B(\rm\it{B^{\rm{0}} \rightarrow K^{{\rm{0}}}\mu^{\pm}e^{\mp}})$ corresponds, via isospin invariance, to $ \cal B(\rm\it{{B^{+} \rightarrow K^{+}\mu^{+}e^{-}}}) + \cal B(\rm\it{{B^{+} \rightarrow K^{+}\mu^{-}e^{+}}})$. 
The significance of the signal yield for $B^{+} \rightarrow K^{+}\mu^{+}e^{-}$ channel is $2.8\sigma$.
To estimate the signal significance for this mode, we have generated a large sample of pseudoexperiments with $N_{\rm sig}^{\rm true}=0$ and estimated the number of cases which have $N_{\rm sig}^{\rm fit}> N_{\rm sig}(\rm{observed})$. The confidence level obtained is then translated to significance.
We calculate the upper limit for these modes at $90\%$ CL using a frequentist method. In this method, for different numbers of signal events $N_{\rm{sig}} (\rm{gen})$, we generate 1000 Monte Carlo experiments with signal and background PDFs, with each set of events being statistically equivalent to our data sample of $711\invfb$. We fit all these simulated data sets, and, for each value of $N_{\rm{sig}}(\rm{gen)}$, we calculate the fraction of MC experiments that have $N_{\rm sig} \leq N_{\rm{sig}}(\rm{data})$. The $90\%$ CL upper limit is taken to be the value of $N_{\rm{sig}}(\rm{gen})$ (called here $N_{\rm sig}^{\rm{UL}}$) for which $10\%$ of the experiments have $N_{\rm{sig}} \leq N_{\rm{sig}}(\rm{data})$. The upper limit on the branching fraction is then derived using the formula:
\begin{center}
$\cal B^{\rm{UL}}$ $= \dfrac{N_{\rm sig}^{\rm{UL}}}{N_{B\bar{B}} \times 2 \times f^{+-(00)} \times \varepsilon}$,
\end{center} 
where $N_{B\bar{B}}$ is the number of $B\bar{B}$ pairs = $(772 \pm 11) \times 10^{6}$, $f^{+-(00)}$ is the branching fraction ${\cal B}[\Y4S \to B^{+}B^{-}]$  $({\cal B}[\Y4S \to B^{0}\bar{B^{0}}])$ for charged (neutral) $B$ decays, and $\varepsilon$ is the signal reconstruction efficiency calculated from signal MC samples. The systematic uncertainty in $\cal B^{\rm{UL}}$ is included by smearing the $N_{\rm sig}$ obtained from the MC fits with the fractional systematic uncertainty (discussed in Section~\ref{Sec:syst}). The results are listed in Table~\ref{LFV_table}. 
\begin{table}[h]
\begin{center}
\caption{Branching fraction UL calculation at $90\%$ CL for LFV $B \rightarrow K\mu e$ decays.}
\label{LFV_table}
\begin{tabular}{l c c c c}
Mode & $\varepsilon~(\%)$ & $N_{\rm sig}$ & $N_{\rm sig}^{\rm{UL}}$ & $\cal B^{\rm{(UL)}}$ $(10^{-8})$ \\ \hline
$B^{+} \rightarrow K^{+}\mu^{+}e^{-}$ & $29.4$&$11.6^{+6.1}_{-5.5}$ & $19.9$ &$8.5$  \\
$B^{+} \rightarrow K^{+}\mu^{-}e^{+}$ & $31.2$&$1.7^{+3.6}_{-2.2}$ & $7.5$ &$3.0$\\
$B^{0} \rightarrow K^{0}\mu^{\pm}e^{\mp}$ & $20.9$&$-3.3^{+4.0}_{-2.8}$ & $3.0$& $3.8$ \\ \hline
\end{tabular}
\end{center}
\end{table}

\section{Systematic uncertainties}
\label{Sec:syst}

Systematic uncertainties arising due to lepton identification is 0.3\% (0.4\%) for each muon (electron) selection. This uncertainty is calculated using an inclusive $J/\psi \rightarrow \ell^{+}\ell^{-}$, $\ell =e$ or $\mu$ sample. 
Uncertainty due to hadron identification is 0.8\% for $K^{+}$ using $D^{*+} \rightarrow
{{D}}^{0}(K^-\pi^+)\pi^+$ sample and 1.6\% for \KS~\cite{ks_syst}. 
The systematic uncertainty due to charged track reconstruction is $0.35\%$ per track estimated by using the partially reconstructed $D^{*-} \rightarrow {\bar{D}}^{0}\pi^{-}$, ${\bar{D}}^{0} \rightarrow \pi^{+}\pi^{-}\KS$, and $\KS \rightarrow \pi^{+}\pi^{-}$ events. 
The uncertainty in efficiency due to limited MC statistics is about $0.2\%$, and the uncertainty in the number of $B\bar{B}$ events is $1.4\%$. 
The systematic uncertainty in the branching fraction ${\cal B}[\Y4S \to B^{+}B^{-}]~({\cal B}[\Y4S \to B^{0}\bar{B^{0}}])$ is 1.2\%~\cite{pdg}. 
We compare the efficiency of the ${\cal O} > {\cal O}_{\rm min}$ criterion between data and MC samples with the control channel $B \rightarrow J/\psi K$, $J/\psi\to\ell^+\ell^-$; the differences between data and MC simulation ($0.9$-$1.2$\%) are corrected and the corresponding uncertainty ($0.2$-$0.3$\%) is assigned as a systematic uncertainty. 
The uncertainty due to  PDF shapes is evaluated by varying the fixed shape parameters by $\pm 1 \sigma$ and repeating the fit; the change in the central value of $N^{}_{\rm sig}$ is taken as the systematic uncertainty, which ranges from 0.1 to 0.6\%. 
The uncertainty due to the fixed yield of continuum events is estimated by varying the yield by $\pm 1 \sigma$ in the fit; the resulting variation in $N_{\rm sig}$ is less than 1\%. 
The charmless $B \rightarrow K\pi^{+}\pi^{-}$ background fixed in the fit for the modes with muon final states is varied within $\pm 1\sigma$ in the fit, and the change in $N_{\rm sig}$ is assigned as systematic, which is $0.1$-$0.2\%$. 
The decay model systematic for $B \rightarrow K\ell^{+}\ell^{-}$ modes is evaluated by comparing reconstruction efficiencies calculated from MC samples generated with different models~\cite{mc_decay_model, mc_decay_model2} and is $0.3$ to $2.0\%$ depending on the $q^2$ bin. For the $B \rightarrow J/\psi K$ branching fraction, we have considered all the sources except for the contribution due to fixed continuum or charmless $B \rightarrow K\pi^{+}\pi^{-}$ events and the decay model. 
The systematic uncertainties such as hadron identification, track reconstruction, number of $B\bar{B}$ events, and the ratio ${\cal B}[\Y4S \to B^{+}B^{-}]~({\cal B}[\Y4S \to B^{0}\bar{B^{0}}])$ cancel out in the double ratio of $R_K(J/\psi)$, while for $A_{I}(J/\psi K)$ the sources that divide out are lepton identification and number of $B\bar{B}$ events as listed in Table~\ref{jpsi_systematic}. In the case of $R_{K}$, systematic uncertainties due to hadron identification, charged track reconstruction, number of $B\bar{B}$ events, and the ${\cal B}[\Y4S \to B^{+}B^{-}]~({\cal B}[\Y4S \to B^{0}\bar{B^{0}}])$ cancel, while for the $A_{I}$ measurement lepton identification and the number of $B\bar{B}$ events cancel.
\begin{table}[htb]
    \centering
    \caption{Relative systematic uncertainties (\%) for ${\cal B}(B \rightarrow J/\psi K)$, $R_K(J/\psi)$, and absolute uncertainty for $A_{I} (B \rightarrow J/\psi K)$.}
    \label{jpsi_systematic}
    \begin{tabular}{l c c c c c}
     Sources & $B^{+} \rightarrow J/\psi K^{+}$ & $B^{0} \rightarrow J/\psi \KS$ & $R_{K^{+}}(J/\psi)$ & $R_{K^{0}}(J/\psi)$& $A_{I} (J/\psi K)$ \\ \hline 
     Lepton identification & $\pm 0.68$& $\pm 0.68$ & $\pm 0.97$ & $\pm 0.97$& $-$\\ 
     Kaon identification & $\pm 0.80$ & $-$ & $-$ & $-$& $\pm 0.007$ \\
     $K^{0}_{S}$ identification & $-$ & $\pm 1.57$ & $-$ & $-$& $\pm 0.002$ \\
     Track reconstruction & $\pm 1.05$ & $\pm 1.40$ & $-$ & $-$& $\pm 0.002$\\
     Efficiency calculation & $\pm 0.14$ & $\pm 0.18$& $\pm 0.20$ & $\pm 0.25$& $\pm 0.001$  \\ 
     Number of $B\bar{B}$ pairs & $\pm 1.40$ & $\pm 1.40$& $-$& $-$& $-$\\
     $f^{+-(00)}$ & $\pm 1.20$ & $\pm 1.20$ & $-$& $-$& $\pm 0.012$ \\
     $\cal O_{\rm min}$ & $\pm 0.16$ & $\pm 0.28$ & $\pm 0.24$ & $\pm 0.39$ & $\pm 0.001$ \\
     PDF shape parameters & $^{+0.15}_{-0.20}$ & $^{+0.05}_{-0.10}$& $^{+0.22}_{-0.31}$ & $^{+0.10}_{-0.20}$&$\pm 0.002$  \\ \hline
     Total & $\pm 2.38$ & $\pm 2.90$ & $^{+1.05}_{-1.07}$ & $^{+1.08}_{-1.09}$& $\pm 0.014$ \\ \hline
    \end{tabular}
\end{table}

\section{Summary}
\label{Sec:con}

In summary, we have measured the differential branching fractions, their ratios ($R_{K}$), and the $C\!P$-averaged isospin asymmetry ($A_{I}$) for the $B\to K \ell^{+}\ell^{-}$ decays as a function of $q^{2}$. 
The branching fractions for $B \rightarrow K \ell^{+}\ell^{-}$ modes are
\begin{center}
$\cal B\rm{\it{(B^{+} \rightarrow K^{+}\ell^{+}\ell^{-})}}$ $= (5.99^{+0.45}_{-0.43}\pm 0.14) \times 10^{-7}$,\\
$\cal B\rm{\it{(B^{\rm{0}} \rightarrow K^{\rm{0}}\ell^{+}\ell^{-})}}$ $= (3.51^{+0.69}_{-0.60}\pm 0.10) \times 10^{-7}$.\\
\end{center}
The branching fractions for $B^{+} \rightarrow J/\psi K^{+}$, and $B^{0} \rightarrow J/\psi K^{0}$ are $(1.032\pm 0.007\pm 0.024) \times~10^{-3}$, and $(0.902 \pm 0.010 \pm 0.026) \times~10^{-3}$, respectively. These are the single most precise measurements to date. The $R_{K}$ values for different $q^{2}$ bins are consistent with the SM predictions, and the value for the whole $q^{2}$ range is $1.10^{+0.16}_{-0.15}\pm 0.02$. The results for five $q^{2}$ bins are
\begin{eqnarray*}
R_{K} =                                                     \begin{cases}                                            1.01~ ^{+0.28}_{-0.25}  \pm 0.02    & q^{2} \in (0.1,4.0)~\mathrm{\
    \,Ge\kern -0.1em V^2\!/}c^4 \, , \\
  0.85~ ^{+0.30}_{-0.24}  \pm 0.01    & q^{2} \in (4.00,8.12)~\mathrm{\
    \,Ge\kern -0.1em V^2\!/}c^4 \, , \\
  1.03~ ^{+0.28}_{-0.24}  \pm 0.01    & q^{2} \in (1.0,6.0)~\mathrm{\
    \,Ge\kern -0.1em V^2\!/}c^4 \, , \\
 1.97~ ^{+1.03}_{-0.89}  \pm 0.02    & q^{2} \in (10.2,12.8)~\mathrm{\
    \,Ge\kern -0.1em V^2\!/}c^4 \, , \\
  1.16~ ^{+0.30}_{-0.27}  \pm 0.01    & q^{2} > 14.18~\mathrm{\
    \,Ge\kern -0.1em V^2\!/}c^4 \ .  \\
\end{cases}
\end{eqnarray*}
Our result of $R_{K^{+}}$ for the bin of interest, $q^{2} \in (1.0,6.0)\qq$, is higher than the LHCb result~\cite{ex:lhcb:rk,lhcb_rk} by 1.6$\sigma$.
The $A_{I}$ values for almost all the bins for different channels show a negative asymmetry. For the bin $q^{2}\in(1.0,6.0)\qq$, the obtained $A_{I}$ value deviates from zero by 2.6$\sigma$ for the mode with muon final states. The $A_{I}$ value for the whole $q^{2}$ range is $-0.19^{+0.07}_{-0.06}\pm 0.01$.
We see no deviation in differential branching fractions for the mode $B^+\to K^+\mu\mu$, where LHCb~\cite{lhcb_dBR} observes lower values than the standard model predictions, though not inconsistent with our result.
The values for this observable are lower than the theoretical prediction for neutral $B$ decays, reflecting $A_{I}<1$. 
We have also searched for the lepton-flavor-violating $B \rightarrow K\mu e$ decays and set upper limits on their branching fractions at 90\% CL:
\begin{center}
$\cal B\rm{\it{(B^{+} \rightarrow K^{+}\mu^{+}e^{-})}}$ $< 8.5 \times 10^{-8}$,\\
$\cal B\rm{\it{(B^{+} \rightarrow K^{+}\mu^{-}e^{+})}}$ $< 3.0 \times 10^{-8}$,\\
$\cal B\rm{\it{(B^{{\rm{0}}} \rightarrow K^{{\rm{0}}}\mu^{\pm}e^{\mp})}}$ $< 3.8 \times 10^{-8}$.\\
\end{center}
We improve the existing limit on the neutral decay mode by an order of magnitude. More precisely, the limit of BaBar~\cite{BaBar_lfv} is $2.7\times 10^{-7}$, $i.e.,$ the improvement is by a factor of $7.1$.

\section{Acknowledgments}
KT wishes to thank S. Descotes-Genon for useful discussions.
We thank the KEKB group for the excellent operation of the
accelerator; the KEK cryogenics group for the efficient
operation of the solenoid; and the KEK computer group, and the Pacific Northwest National
Laboratory (PNNL) Environmental Molecular Sciences Laboratory (EMSL)
computing group for strong computing support; and the National
Institute of Informatics, and Science Information NETwork 5 (SINET5) for
valuable network support.  We acknowledge support from
the Ministry of Education, Culture, Sports, Science, and
Technology (MEXT) of Japan, the Japan Society for the 
Promotion of Science (JSPS), and the Tau-Lepton Physics 
Research Center of Nagoya University; 
the Australian Research Council including grants
DP180102629, % Sevior
DP170102389, % Varvell
DP170102204, % Yabsley
DP150103061, % Urquijo
FT130100303; % Urquijo;
Austrian Science Fund (FWF);
the National Natural Science Foundation of China under Contracts
No.~11435013,  %Zhen-An Liu
No.~11475187,  %Chang-Zheng Yuan
No.~11521505,  %Chang-Zheng Yuan
No.~11575017,  %Cheng-Ping Shen
No.~11675166,  %Wen-Biao Yan
No.~11705209;  %Yi-Ming Li
Key Research Program of Frontier Sciences, Chinese Academy of Sciences (CAS), Grant No.~QYZDJ-SSW-SLH011; % Chang-Zheng Yuan
the  CAS Center for Excellence in Particle Physics (CCEPP); %Chang-Zheng Yuan, …
the Shanghai Pujiang Program under Grant No.~18PJ1401000;  %Tao Luo
the Ministry of Education, Youth and Sports of the Czech
Republic under Contract No.~LTT17020;
the Carl Zeiss Foundation, the Deutsche Forschungsgemeinschaft, the
Excellence Cluster Universe, and the VolkswagenStiftung;
the Department of Science and Technology of India; 
the Istituto Nazionale di Fisica Nucleare of Italy; 
National Research Foundation (NRF) of Korea Grants
No.~2016R1D1A1B01010135,  No.~2016R1D1A1B02012900,  No.~2018R1A2B3003643, No.~2018R1A6A1A06024970, No.~2018R1D1A1B07047294, No.~2019K1A3A7A09033840;
Radiation Science Research Institute, Foreign Large-size Research Facility Application Supporting project, the Global Science Experimental Data Hub Center of the Korea Institute of Science and Technology Information and KREONET/GLORIAD;
the Polish Ministry of Science and Higher Education and 
the National Science Center;
the Grant of the Russian Federation Government, Agreement No.~14.W03.31.0026; % from 15.02.2018;
the Slovenian Research Agency;
Ikerbasque, Basque Foundation for Science, Spain;
the Swiss National Science Foundation; 
the Ministry of Education and the Ministry of Science and Technology of Taiwan;
and the United States Department of Energy and the National Science Foundation.


\begin{thebibliography}{99}
\bibitem{cc}
  The inclusion of the charge-conjugate decay mode is implied unless otherwise stated.
\end{thebibliography}
\end{document}